\newif\ifreport
\newif\ifnotreport
\reporttrue 

\ifreport
    \notreportfalse
\else
    \notreporttrue
\fi

\documentclass[a4paper]{article}
\ifreport
    
\usepackage{vmargin}    

\usepackage{postscript} 
\usepackage{pslatex}    

\usepackage[cspex,bbgreekl]{mathbbol} 
\usepackage{pifont}     
\usepackage{eurosym}    
\usepackage{amssymb}    
\usepackage{amstext}    
\usepackage{lgrind}     
\usepackage{algorithm}  
\usepackage{algorithmic}
\usepackage{booktabs}   

\usepackage[bf,sf,medium,compact]{titlesec} 
\usepackage{fancyhdr}                

\newif\iftth

\setmarginsrb{1.5cm}{1.5cm}{1.5cm}{1.5cm}{12pt}{25pt}{12pt}{30pt}

\renewenvironment{itemize}%
  {\begin{list}{$\bullet$}%
               {\setlength{\parsep}{0pt}
                \setlength{\itemsep}{0pt}}}%
  {\end{list}}

\newcommand{\vb}[1]{\verb+#1+}

\fi

\usepackage{amsfonts}
\usepackage{amsmath}
\usepackage{mathptmx}
\usepackage{helvet}
\usepackage{courier}
\usepackage{makeidx}
\usepackage{multicol}
\usepackage{graphicx}
\usepackage{xspace}
\usepackage[T1]{fontenc}
\usepackage{algorithm}
\usepackage{algorithmic}
\usepackage{listings}
\usepackage{longtable}
\usepackage[square,sort,comma,numbers]{natbib}
\usepackage{bibentry}
\ifnotreport
    \nobibliography*
\fi

\ifreport
    \usepackage{hyperref}
\fi

\begin{document}
\newcommand{\DEVS}{\textsf{DEVS}\xspace}
\newcommand{\PDEVS}{\textsf{Parallel DEVS}\xspace}
\newcommand{\CELLDEVS}{\textsf{Cell-DEVS}\xspace}
\newcommand{\CELL}{\textsf{Cellular Automata}\xspace}
\newcommand{\DSDEVS}{\textsf{Dynamic Structure DEVS}\xspace}
\newcommand{\CDEVS}{\textsf{Classic DEVS}\xspace}
\newcommand{\DES}{\textsf{DES}\xspace}
\newcommand{\SC}{\textsf{Statecharts}\xspace}

\newcommand{\SGREEN}{\textsc{green}\xspace}
\newcommand{\SYELLOW}{\textsc{yellow}\xspace}
\newcommand{\SRED}{\textsc{red}\xspace}
\newcommand{\SMANUAL}{\textsc{manual}\xspace}
\newcommand{\STOMANUAL}{\textsc{going\_manual}\xspace}
\newcommand{\STOAUTO}{\textsc{going\_auto}\xspace}

\newcommand{\EGREEN}{\textit{show\_green}\xspace}
\newcommand{\EYELLOW}{\textit{show\_yellow}\xspace}
\newcommand{\ERED}{\textit{show\_red}\xspace}
\newcommand{\EMANUAL}{\textit{toManual}\xspace}
\newcommand{\EAUTO}{\textit{toAuto}\xspace}
\newcommand{\ETAKEBREAK}{\textit{take\_break}\xspace}
\newcommand{\EGOTOWORK}{\textit{go\_to\_work}\xspace}
\newcommand{\EOFF}{\textit{turn\_off}\xspace}

\newcommand{\DELAYGREEN}{\textit{$\mathit{delay}_\mathit{green}$}}
\newcommand{\DELAYYELLOW}{\textit{$\mathit{delay}_\mathit{yellow}$}}
\newcommand{\DELAYRED}{\textit{$\mathit{delay}_\mathit{red}$}}

\ifnotreport
	\newcommand{\reflection}[1]{
		{
			\vbox{
			\setlength{\parindent}{0cm}
			\textsc{Reflection Question}

			\textit{#1}
			}
		}
	}
\fi

\ifreport
    \title{An Introduction to Classic DEVS}
\else
    \title{Discrete-Event Modelling for Queueing Systems and Performance Analysis}
\fi

\ifreport
    \newcommand{\footremember}[2]{%
        \footnote{#2}
        \newcounter{#1}
        \setcounter{#1}{\value{footnote}}%
    }
    \newcommand{\footrecall}[1]{%
        \footnotemark[\value{#1}]%
    } 

    \author{
        \begin{tabular}{cc}
            Yentl Van Tendeloo\footremember{UA}{University of Antwerp, Belgium}  &   Hans Vangheluwe\footrecall{UA} \footnote{Flanders Make, Belgium} \footnote{McGill University, Montr\'{e}al, Canada}\\
        \end{tabular} \\
        \{Yentl.VanTendeloo,Hans.Vangheluwe\}@uantwerpen.be
    }
\else
    \author{Yentl Van Tendeloo \and Hans Vangheluwe}
    \institute{Yentl Van Tendeloo \at University of Antwerp, Belgium \\ \email{Yentl.VanTendeloo@uantwerpen.be}
    \and Hans Vangheluwe \at University of Antwerp - Flanders Make, Belgium; McGill University, Canada \\ \email{Hans.Vangheluwe@uantwerpen.be}}
\fi

\maketitle

\ifreport
    \begin{abstract}
        DEVS is a popular formalism for modelling complex dynamic systems using a
discrete-event abstraction. At this abstraction level, a timed sequence of
pertinent ``events'' input to a system (or internal, in the case of timeouts)
cause instantaneous changes to the state of the system. Between events, the
state does not change, resulting in a piecewise constant state trajectory.
Main advantages of DEVS are its rigorous formal definition, and its support for
modular composition.

This chapter introduces the Classic DEVS formalism in a bottom-up fashion, 
using a simple traffic light example. The syntax and operational semantics 
of Atomic (i.e., non-hierarchical) models are introduced first.
The semantics of Coupled (hierarchical) models is then given by translation into
Atomic DEVS models. As this formal ``flattening'' is not efficient,
a modular abstract simulator which operates directly on the coupled model is also 
presented. This is the common basis for subsequent efficient implementations.
We continue to actual applications of DEVS modelling and simulation, as seen in
performance analysis for queueing systems. 
Finally, we present some of the shortcomings in the Classic DEVS formalism, and
show solutions to them in the form of variants of the original formalism.

    \end{abstract}
\else
    \abstract{DEVS is a popular formalism for modelling complex dynamic systems using a
discrete-event abstraction. At this abstraction level, a timed sequence of
pertinent ``events'' input to a system (or internal, in the case of timeouts)
cause instantaneous changes to the state of the system. Between events, the
state does not change, resulting in a piecewise constant state trajectory.
Main advantages of DEVS are its rigorous formal definition, and its support for
modular composition.

This chapter introduces the Classic DEVS formalism in a bottom-up fashion, 
using a simple traffic light example. The syntax and operational semantics 
of Atomic (i.e., non-hierarchical) models are introduced first.
The semantics of Coupled (hierarchical) models is then given by translation into
Atomic DEVS models. As this formal ``flattening'' is not efficient,
a modular abstract simulator which operates directly on the coupled model is also 
presented. This is the common basis for subsequent efficient implementations.
We continue to actual applications of DEVS modelling and simulation, as seen in
performance analysis for queueing systems. 
Finally, we present some of the shortcomings in the Classic DEVS formalism, and
show solutions to them in the form of variants of the original formalism.
}
\fi

\ifnotreport
\vfill

    \vbox{
    \textsc{Learning Objectives}

    After completing this chapter we expect you to be able to:
    \begin{itemize}
    \item Understand the difference between \DEVS and other (similar) formalisms
    \item Explain the semantics of a given \DEVS model
    \item Understand the relation and difference between a \DEVS model and its simulator
    \item Apply \DEVS to a simple queueing problem
    \item Understand the major shortcomings of \DEVS and their proposed solutions
    \end{itemize}
    }
\fi

\section{Introduction}
\DEVS~\cite{ClassicDEVS} is a popular formalism for modelling complex dynamic systems using a discrete-event abstraction.
At this abstraction level, a timed sequence of pertinent ``events'' input to a system cause instantaneous changes to the state of the system.
These events can be generated externally (\textit{i.e.}, by another model) or internally (\textit{i.e.}, by the model itself due to timeouts).
The next state of the system is defined based on the previous state of the system and the event.
Between events, the state does not change, resulting in a piecewise constant state trajectory.
Simulation kernels must only consider states at which events occur, skipping over all intermediate points in time.
This is in contrast with discrete time models, where time is incremented with a fixed increment, and only at these times is the state updated.
Discrete event models have the advantage that their time granularity can become (theoretically) unbounded, whereas time granularity is fixed in discrete time models.
Nonetheless, the added complexity makes it unsuited for systems that naturally have a fixed time step.

Main advantages of \DEVS compared to other discrete event formalisms are its rigorours formal definition, and its support for modular composition.
Comparable discrete event formalisms are \DES and \SC, though significant differences exist.

Compared to \DES, \DEVS offers modularity which makes it possible to nest models inside of components, thus generating a hierarchy of models.
This hierarchy necessitates couplings and (optionally) ports, which are used for all communication between two components.
In contrast, \DES models can directly access other models and send them events in the form of method invocation.
Additionaly, \DEVS offers a cleaner split between the simulation model and the simulation kernel.

Compared to \SC, \DEVS offers a more rigorous formal definition and a different kind of modularity.
\SC leaves a lot of room for interpretation, resulting in a wide variety of interpretations~\cite{Statemate,Rhapsody}.
In contrast, \DEVS is completely formally defined, and there is a reference algorithm (\textit{i.e.}, an abstract simulator).
While both \DEVS and \SC are modular formalisms, \SC creates hierarchies through composite states, whereas \DEVS uses composite models for this purpose.
Both have their distinct advantages, making both variants useful in practice.

This chapter provides an introductory text to \DEVS (often referred to as \CDEVS nowadays) through the use of a simple example model in the domain of traffic lights.
We start from a simple autonomous traffic light, which is incrementally extended up to a traffic light with policeman interaction.
Each increment serves to introduce a new component of the \DEVS formalism and the corresponding (informal) semantics.
We start with atomic (\textit{i.e.}, non-hierarchical) models in Section~\ref{sec:atomic}, and introduce coupled (\textit{i.e.}, hierarchical) models in Section~\ref{sec:coupled}.
An abstract simulator, defining the reference algorithm, is covered in Section~\ref{sec:semantics}.
Section~\ref{sec:applications} moves away from the traffic light example and presents a model of a simple queueing system.
Even though \DEVS certainly has its applications, several variants have spawned to tackle some of its shortcomings.
These variants, together with a rationale and the differences, are discussed in Section~\ref{sec:variants}.
Finally, Section~\ref{sec:conclusions} summarizes the chapter.

\section{Atomic DEVS models}
\label{sec:atomic}
We commence our explanation of \DEVS with the atomic models.
As their name suggests, atomic models are the indivisible building blocks of a model.

Throughout this section, we build up the complete formal specification of an atomic model, introducing new concepts as they become required.
In each intermediate step, we show and explain the concepts we introduce, how they are present in the running example, and how this influences the semantics.
The domain we will use as a running example throughout this chapter is a simple traffic light.

\subsection{Autonomous Model}
\begin{figure}[t]
    \begin{minipage}{0.60\textwidth}
        \center
        \includegraphics[width=\textwidth]{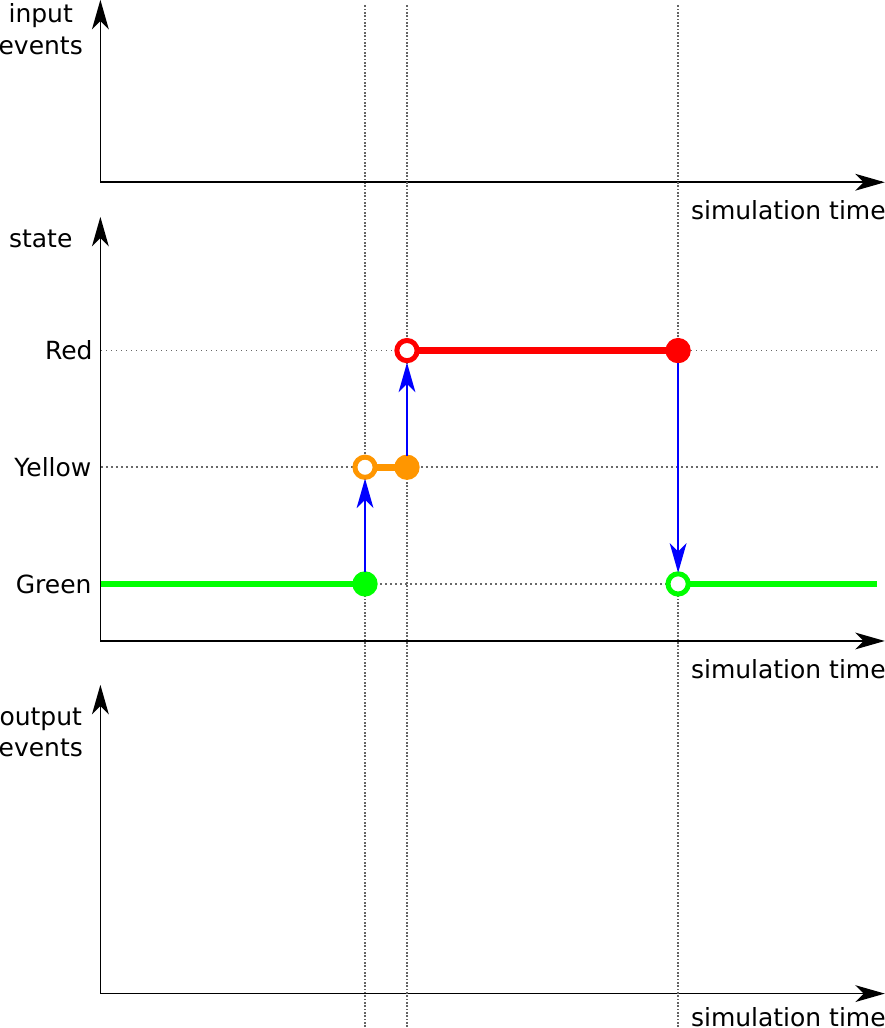}
        \caption{Trace of the autonomous traffic light.}
        \label{fig:trafficlight_autonomous_trace}
    \end{minipage}
    \hfill
    \begin{minipage}{0.35\textwidth}
        \center
        \includegraphics[width=0.4\textwidth]{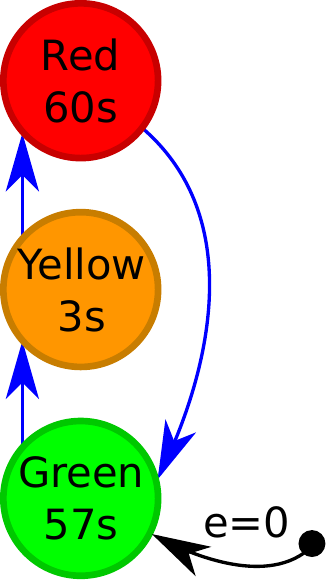}
        \caption{Model generating trace in Figure~\ref{fig:trafficlight_autonomous_trace}.}
        \label{fig:trafficlight_autonomous_model}
    \end{minipage}
\end{figure}

The simplest form of a traffic light is an autonomous traffic light.
Looking at it from the outside, we expect to see a trace similar to that of Figure~\ref{fig:trafficlight_autonomous_trace}.
Visually, Figure~\ref{fig:trafficlight_autonomous_model} presents an intuitive representation of a model that could generate this kind of trace.

Trying to formally describe Figure~\ref{fig:trafficlight_autonomous_model}, we distinguish these elements:
\begin{enumerate}
    \item \textbf{State Set ($S$)}

        The most obvious aspect of the traffic light is the state it is in, which is indicated by the three different colours it can have.
        These states are \textit{sequential}: the traffic light can only be in one of these states at the same time\footnote{In contrast to, say, Statecharts.}.
        The set of states is not limited to enumeration style as presented here, but can contain an arbitrary number of attributes.
        \[
            S : \times^n_{i=1} S_i
        \]

    \item \textbf{Time Advance ($ta$)}

        For each of the states just defined, we notice the timeout in them.
        Clearly, some states take longer to process than others.
        For example, whereas we will stay in green and red a relatively long time, the time in the yellow state is only brief.
        This function needs to be defined for each and every element of the state set, and needs to deterministically return a duration.
        The duration can be any positive real number, including zero and infinity.
        A negative time is disallowed, as this would require simulation to go back in time.
        \DEVS allows a time advance of exactly zero, even though this is impossible in real life.
        Two use cases for this exist: the delay might be very small and irrelevant to the problem we are modelling, or the state is an artificial state, without any real-world equivalent (\textit{e.g.}, as part of a design pattern).
        Note that \DEVS does not consider time bases, despite the use of seconds in our visualization.
        Simulation time is just a real number, and the interpretation given to it is up to the user.
        Whether these units indicate seconds, years, or even $\pi$ seconds, is completely up to the users, as long as it is fixed throughout the simulation.
        \[
            ta : S \to \mathbb{R}^+_{0,+\infty}
        \]

    \item \textbf{Internal Transition ($\delta_{int}$)}

        With the states and timeouts defined, the final part is the definition of which is the next state from a given state.
        This is the job of the internal transition function, which gives the next state for each and every state.
        As it is a function, every state has at most one next state, preventing any possible ambiguity.
        Note that the function does not necessarily have to be total, nor injective: some states might not have a next state (\textit{i.e.}, if the time advance was specified as $+\infty$), and some states have the same state as next state.
        Up to now, only the internal transition function is described as changing the state.
        Therefore, it is not allowed for other functions (\textit{e.g.}, time advance) to modify the state: their state access is read-only.
        \[
            \delta_{int} : S \to S
        \]

    \item \textbf{Initial Total State ($q_{init}$)}
        
        We also need to define the initial state of the system.
        While this is not present in the original specification of the DEVS formalism, we include it here as it is a vital element of the model~\cite{InitializedDEVS}.
        But note that, instead of being an ``initial state ($s_{init}$)'', it is a total state.
        This means that we not only select the initial state of the system, but also define how long we are already in this state.
        Elapsed time is therefore added to the definition of the initial total state, to allow more flexibility when modelling a system.
        To the simulator, it will seem as if the model has already been in the initial state for some time.
        \[
            q_{init} : {(s, e) | s \in S, 0 \leq e \leq ta(s)}
        \]
\end{enumerate}

We describe the model in Figure~\ref{fig:trafficlight_autonomous_model} as a 4-tuple of these three elements.
\begin{align*}
\langle S, q_{init}, \delta_{int}, ta \rangle
\end{align*}
\begin{align*}
S               = \{&\SGREEN, \SYELLOW, \SRED\} \\
q_{init}        = (&\SGREEN, 0.0) \\
\delta_{int}    = \{&\SGREEN \rightarrow \SYELLOW, \\
                    &\SYELLOW \rightarrow \SRED, \\
                    &\SRED \rightarrow \SGREEN\} \\
ta              = \{&\SGREEN \rightarrow \DELAYGREEN, \\
                    &\SYELLOW \rightarrow \DELAYYELLOW, \\
                    &\SRED \rightarrow \DELAYRED\}
\end{align*}

For this simple formalism, we define the semantics as in Algorithm~\ref{algorithm:autonomous}.
The model is initialized with simulation time set to 0, and the state set to the initial state (\textit{e.g.}, \SGREEN).
Simulation updates the time with the return value of the time advance function, and executes the internal transition function on the current state to get the new state.

\begin{algorithm*}
\begin{algorithmic}
    \STATE $time \leftarrow 0$
    \STATE $current\_state \leftarrow initial\_state$
    \STATE $last\_time \leftarrow -initial\_elapsed$
    \WHILE {not termination\_condition()}
        \STATE $time \leftarrow last\_time + ta(current\_state)$
        \STATE $current\_state \leftarrow \delta_{int}(current\_state)$
        \STATE $last\_time \leftarrow time$
    \ENDWHILE
\end{algorithmic}
\caption{\DEVS simulation pseudo-code for autonomous models.}
\label{algorithm:autonomous}
\end{algorithm*}

\subsection{Autonomous Model With Output}
Recall that \DEVS is a modular formalism, with only the atomic model having access to its internal state.
This naturally raises a problem for our traffic light: others have no way of knowing its current state (\textit{i.e.}, its colour).

\begin{figure}[t]
    \center
    \begin{minipage}{0.6\textwidth}
        \center
        \includegraphics[width=\textwidth]{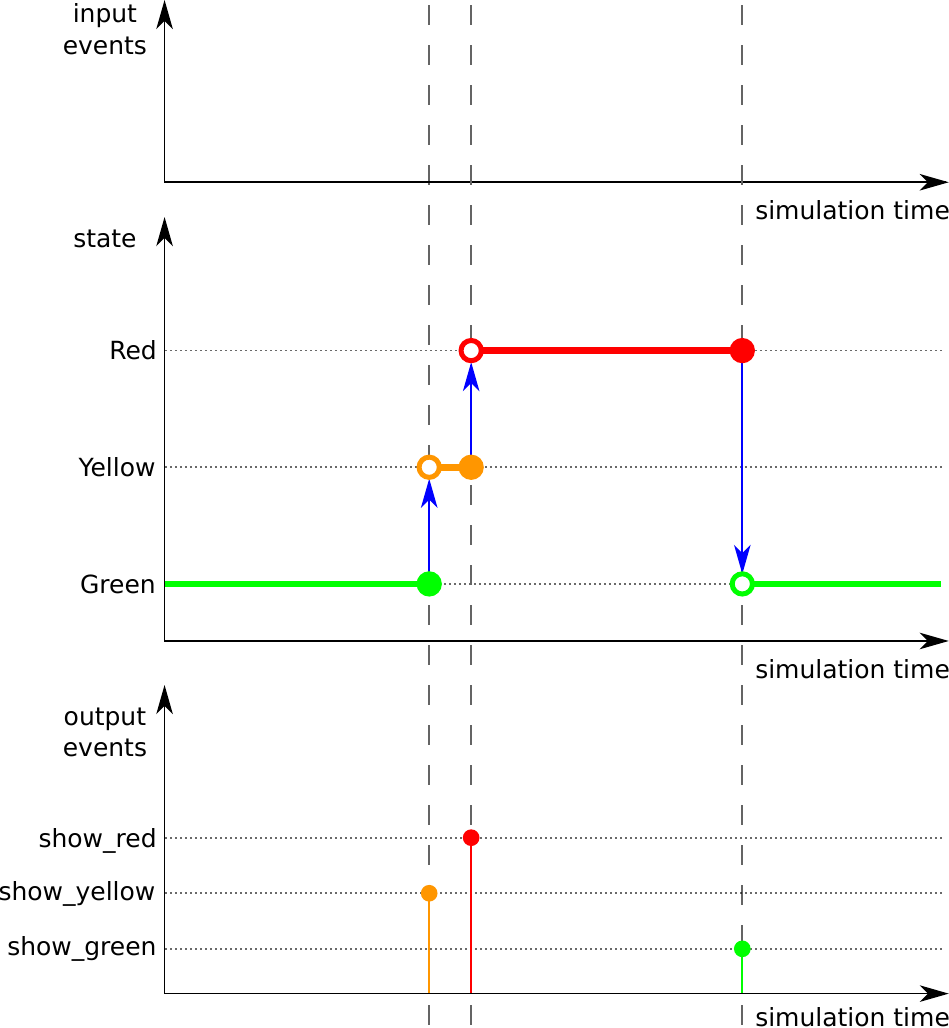}
        \caption{Trace of the autonomous traffic light with output.}
        \label{fig:trafficlight_autonomous_output_trace}
    \end{minipage}
    \hfill
    \begin{minipage}{0.35\textwidth}
        \center
        \includegraphics[width=\textwidth]{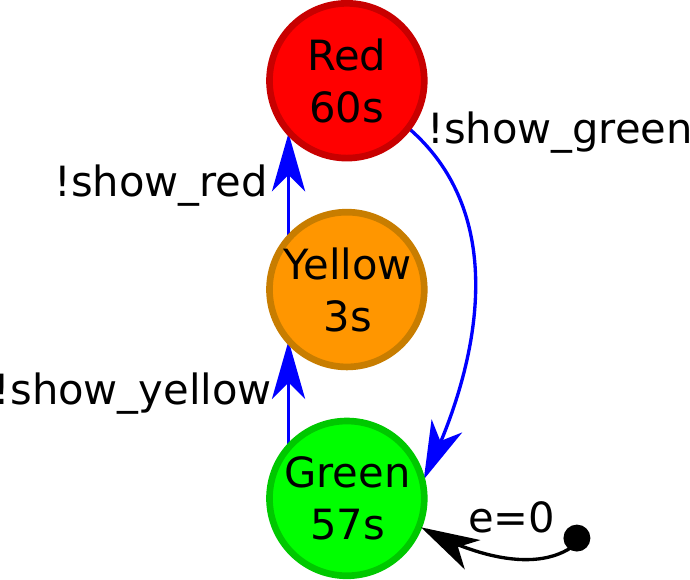}
        \caption{Model generating trace in Figure~\ref{fig:trafficlight_autonomous_output_trace}.}
        \label{fig:trafficlight_autonomous_output_model}
    \end{minipage}
\end{figure}

We therefore want the traffic light to output its colour, in this case in the form of a string (and not as the element of an enumeration).
For now, the output is tightly linked to the set of state, but this does not need to be the case: the possible values to output can be completely distinct from the set of states.
Our desired trace is shown in Figure~\ref{fig:trafficlight_autonomous_output_trace}.
We see that we now output events indicating the start of the specified period.
Recall, also, that \DEVS is a discrete event formalism: the output is only a single event indicating the time and is not a continuous signal.
The receiver of the event thus would have to store the event to know the current state of the traffic light at any given point in time.
Visually, the model is updated to Figure~\ref{fig:trafficlight_autonomous_output_model}, using the exclamation mark on a transition to indicate output generation.

Analysing the updated model, we see that two more concepts are required to allow for output.
\begin{enumerate}
    \item \textbf{Output Set ($Y$)}

        Similarly to defining the set of allowable states, we should also define the set of allowable outputs.
        This set serves as an interface to other components, defining the events they expect to receive.
        Events can have complex attributes as well, though we again limit ourself to simple events for now.
        If ports are used, each port has its own output set.
        \[
            Y : \times^l_{i=1} Y_i
        \]

    \item \textbf{Output Function ($\lambda$)}
        
        With the set of allowable events defined, we still need a function to actually generate the events.
        Similar to the other functions, the output function is defined on the state, and deterministically returns an event (or no event).
        As seen in the Figure of the model, the event is generated \textit{before} the new state is reached.
        This means that instead of the new state, the output function still uses the old state (\textit{i.e.}, the one that is being left).
        For this reason, the output function needs to be invoked right before the internal transition function.
        In the case of our traffic light, the output function needs to return the name of the \textit{next} state, instead of the current state.
        For example, if the output function receives the \SGREEN state as input, it needs to generate a \EYELLOW event.

        Similar to the time advance function, this function does not output a new state, and therefore state access is read-only.
        This might require some workarounds: outputting an event often has some repercussions on the model state, such as removing the event from a queue or increasing a counter.
        Since the state cannot be written to, these changes need to be remembered and executed as soon as the internal transition is executed.

        Note that it is possible for the output function not to return any output, in which case it returns $\phi$.
        \[
            \lambda : S \to Y \cup \{\phi\}
        \]

\end{enumerate}

The model can be described as a 6-tuple.
\begin{align*}
    \langle Y, S, q_{init}, \delta_{int}, \lambda, ta \rangle
\end{align*}
\begin{align*}
Y               = \{&\EGREEN, \EYELLOW, \ERED\} \\
S               = \{&\SGREEN, \SYELLOW, \SRED\} \\
q_{init}        = (&\SGREEN, 0.0) \\
\delta_{int}    = \{&\SGREEN \rightarrow \SYELLOW, \\
                    &\SYELLOW \rightarrow \SRED, \\
                    &\SRED \rightarrow \SGREEN\} \\
\lambda         = \{&\SGREEN \rightarrow \EYELLOW, \\
                    &\SYELLOW \rightarrow \ERED, \\
                    &\SRED \rightarrow \EGREEN\} \\
ta              = \{&\SGREEN \rightarrow \DELAYGREEN, \\
                    &\SYELLOW \rightarrow \DELAYYELLOW, \\
                    &\SRED \rightarrow \DELAYRED\}
\end{align*}

The pseudo-code is slightly altered to include output generation, as shown in Algorithm~\ref{algorithm:autonomous_output}.
Recall that output is generated before the internal transition is executed, so the method invocation happens right before the transition.

\begin{algorithm*}
\begin{algorithmic}
    \STATE $time \leftarrow 0$
    \STATE $current\_state \leftarrow initial\_state$
    \STATE $last\_time \leftarrow -initial\_elapsed$
    \WHILE {not termination\_condition()}
        \STATE $time \leftarrow last\_time + ta(current\_state)$
        \STATE $output(\lambda(current\_state))$
        \STATE $current\_state \leftarrow \delta_{int}(current\_state)$
        \STATE $last\_time \leftarrow time$
    \ENDWHILE
\end{algorithmic}
\caption{\DEVS simulation pseudo-code for autonomous models with output.}
\label{algorithm:autonomous_output}
\end{algorithm*}

\subsection{Interruptable Model}
\begin{figure}[t]
    \center
    \begin{minipage}{0.6\textwidth}
        \center
        \includegraphics[width=\textwidth]{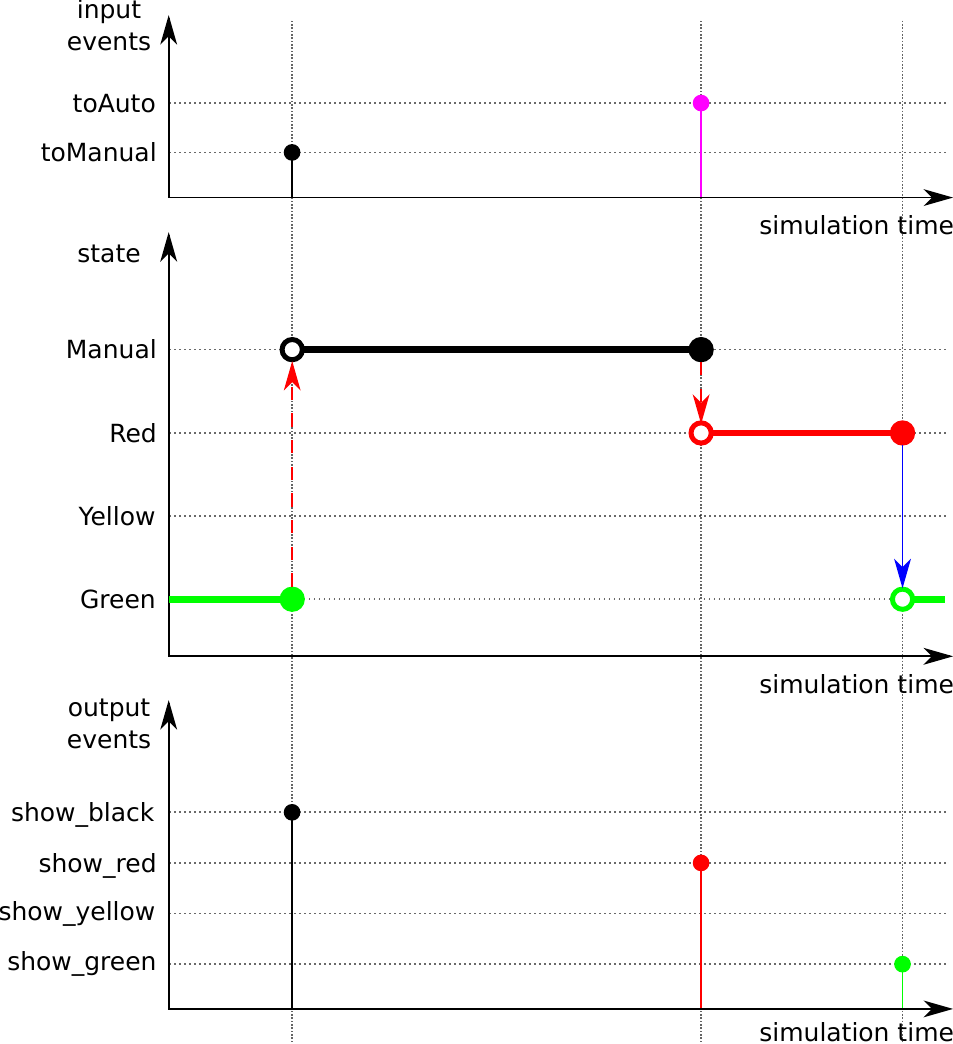}
        \caption{Trace of the autonomous traffic light.}
        \label{fig:trafficlight_interrupt_naive_trace}
    \end{minipage}
    \hfill
    \begin{minipage}{0.35\textwidth}
        \center
        \includegraphics[width=\textwidth]{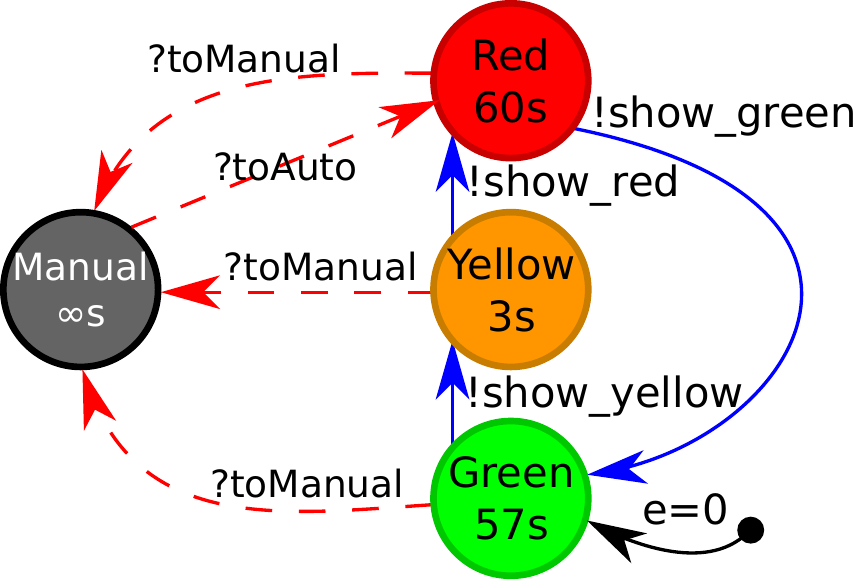}
        \caption{Naive model that should generate the trace in Figure~\ref{fig:trafficlight_interrupt_naive_trace} (but does not).}
        \label{fig:trafficlight_interrupt_naive_model}
    \end{minipage}
\end{figure}

Our current traffic light specification is still completely autonomous.
While this is fine in most circumstances, police might want to temporarily shut down the traffic lights, when they are managing traffic manually.
To allow for this, our traffic light must process externally generated incoming events; such as events from a policeman to shutdown or startup again.
Figure~\ref{fig:trafficlight_interrupt_naive_trace} shows the trace we wish to obtain.
A model generating this trace is shown in Figure~\ref{fig:trafficlight_interrupt_naive_model}, using a question mark to indicate event reception.

We once more require two additional elements in the \DEVS specification.
\begin{enumerate}
    \item \textbf{Input Set ($X$)}

        Similar to the output set, we need to define the events we expect to receive.
        This is again a definition of the interface, such that others know which events are understood by this model.
        \[
            X = \times^m_{i=1} X_i
        \]

    \item \textbf{External Transition ($\delta_{ext}$)}

        Similar to the internal transition function, the external transition function is allowed to define the new state as well.
        First and foremost, the external transition function is still dependent on the current state, just like the internal transition function.
        The external transition function has access to two more values; elapsed time and the input event.
        The \textit{elapsed time} indicates how long it has been for this atomic model since the last transition (either internal or external).
        Whereas this number was implicitly known in the internal transition function (\textit{i.e.}, the value of the time advance function), here it needs to be passed explicitly.
        Elapsed time is a number in the range $[0, ta(s)]$, with $s$ being the current state of the model.
        Note that it is inclusive of both $0$ and $ta(s)$: it is possible to receive an event exactly after a transition happened, or exactly before an internal transition happens.
        The combination of the current state and the elapsed time is often called the \textit{total state} ($Q$) of the model.
        We have previously seen the total state, in the context of the initial total state.
        The received event is the final parameter to this function.
        A new state is deterministically defined through the combination of these three parameters.
        Since the external transition function takes multiple parameters, multiple external transitions might be defined for a single state.
        \begin{gather*}
            \delta_{ext} : Q \times X \to S \\
            Q = \{(s, e) | s \in S, 0 \leq e \leq ta(s)\}
        \end{gather*}
\end{enumerate}

\begin{figure}[t]
    \center
    \begin{minipage}{0.6\textwidth}
        \center
        \includegraphics[width=\textwidth]{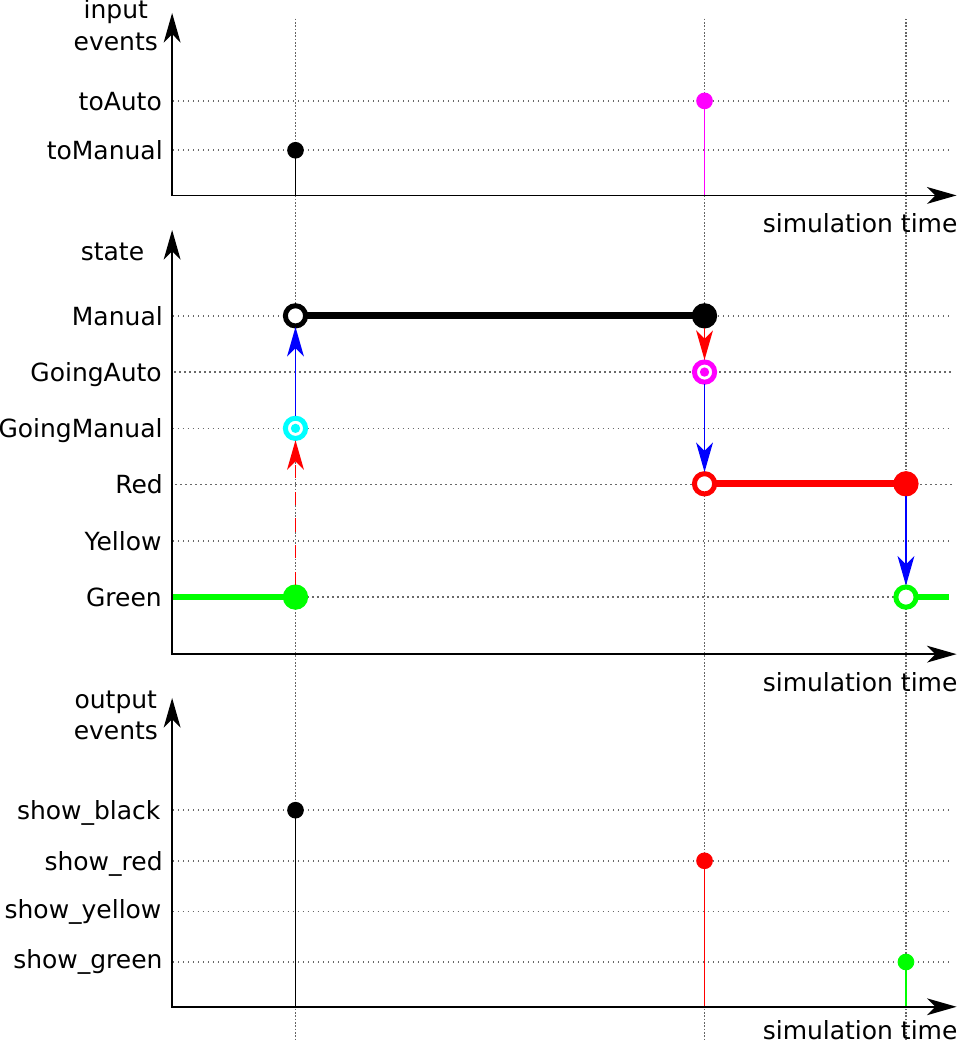}
        \caption{Trace of the interrupt traffic light with corrected artificial states.}
        \label{fig:trafficlight_interrupt_fixed_trace}
    \end{minipage}
    \hfill
    \begin{minipage}{0.35\textwidth}
        \center
        \includegraphics[width=\textwidth]{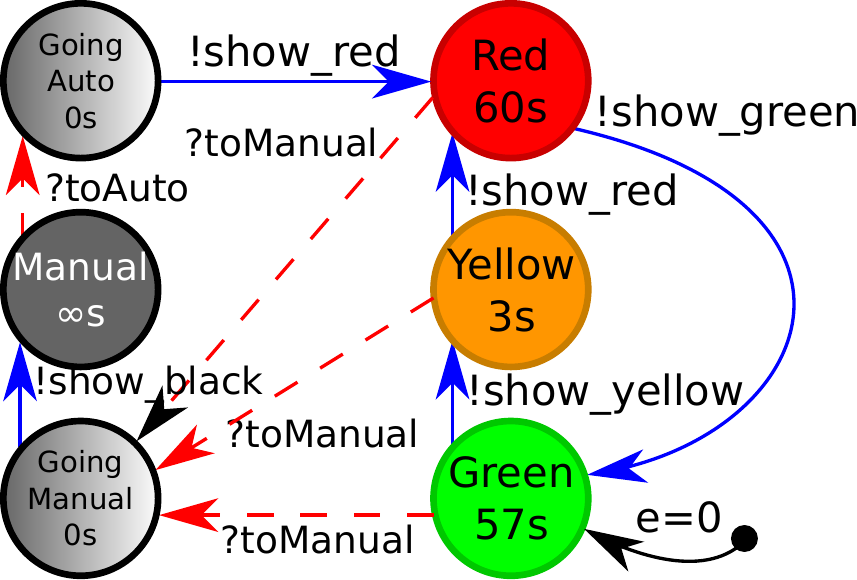}
        \caption{Model generating trace in Figure~\ref{fig:trafficlight_interrupt_fixed_trace}.}
        \label{fig:trafficlight_interrupt_fixed_model}
    \end{minipage}
\end{figure}

While we now have all elements of the \DEVS specification for atomic models, we are not done yet.
When we include the additional state \SMANUAL, we also need to send out an output message indicating that the traffic light is off.
But recall that an output function was only invoked before an internal transition function, so not before an external transition function.
To have an output nonetheless, we need to make sure that an internal transition happens before we actually reach the \SMANUAL state.
This can be done through the introduction of an artificial intermediate state, which times out immediately, and sends out the \EOFF event.
Instead of going to \SMANUAL upon reception of the \EMANUAL event, we go to the artificial state \STOMANUAL.
The time advance of this state is set to 0, since it is only an artificial state without any meaning in the domain under study.
Its output function will be triggered immediately due to the time advance of zero, and the \EOFF output is generated while transferring to \SMANUAL.
Similarly, when we receive the \EAUTO event, we need to go to an artificial \STOAUTO state to generate the \ERED event.
A visualization of the corrected trace and corresponding model is shown in Figure~\ref{fig:trafficlight_interrupt_fixed_trace} and Figure~\ref{fig:trafficlight_interrupt_fixed_model} respectively.

Finally, we give the full specification of the traffic light as an atomic \DEVS model, defined by a 8-tuple.
\begin{align*}
    \langle X, Y, S, q_{init}, \delta_{int}, \delta_{ext}, \lambda, ta \rangle
\end{align*}
\begin{align*}
X               = \{&\EAUTO, \EMANUAL\} \\
Y               = \{&\EGREEN, \EYELLOW, \ERED, \EOFF\} \\
S               = \{&\SGREEN, \SYELLOW, \SRED, \STOMANUAL, \STOAUTO, \SMANUAL\} \\
q_{init}        = (&\SGREEN, 0.0) \\
\delta_{int}    = \{&\SGREEN \rightarrow \SYELLOW, \\
                    &\SYELLOW \rightarrow \SRED, \\
                    &\SRED \rightarrow \SGREEN, \\
                    &\STOMANUAL \rightarrow \SMANUAL, \\
                    &\STOAUTO \rightarrow \SRED\} \\
\delta_{ext}    = \{&(\SGREEN, \_, \EMANUAL) \rightarrow \STOMANUAL \\
                    &(\SYELLOW, \_, \EMANUAL) \rightarrow \STOMANUAL \\
                    &(\SRED, \_, \EMANUAL) \rightarrow \STOMANUAL \\
                    &(\SMANUAL, \_, \EAUTO) \rightarrow \STOAUTO\} \\
\lambda         = \{&\SGREEN \rightarrow \EYELLOW, \\
                    &\SYELLOW \rightarrow \ERED, \\
                    &\SRED \rightarrow \EGREEN, \\
                    &\STOMANUAL \rightarrow \EOFF, \\
                    &\STOAUTO \rightarrow \ERED\} \\
ta              = \{&\SGREEN \rightarrow \DELAYGREEN, \\
                    &\SYELLOW \rightarrow \DELAYYELLOW, \\
                    &\SRED \rightarrow \DELAYRED, \\
                    &\SMANUAL \rightarrow +\infty, \\
                    &\STOMANUAL \rightarrow 0, \\
                    &\STOAUTO \rightarrow 0\}
\end{align*}

Algorithm~\ref{algorithm:interruptable} presents the complete semantics of an atomic model in pseudo-code.
Similar to before, we still have the same simulation loop, but now we can be interrupted externally.
At each time step, we need to determine whether an external interrupt is scheduled before the internal interrupt.
If that is not the case, we simply continue like before, by executing the internal transition.
If there is an external event that must go first, we execute the external transition.

\begin{algorithm*}
\begin{algorithmic}
    \STATE $time \leftarrow 0$
    \STATE $current\_state \leftarrow initial\_state$
    \STATE $last\_time \leftarrow -initial\_elapsed$
    \WHILE {not termination\_condition()}
        \STATE $next\_time \leftarrow last\_time + ta(current\_state)$
        \IF {$time\_next\_event \leq next\_time$}
            \STATE $elapsed \leftarrow time\_next\_event - last\_time$
            \STATE $current\_state \leftarrow \delta_{ext}((current\_state, elapsed), next\_event)$
            \STATE $time \leftarrow time\_next\_event$
        \ELSE
            \STATE $time \leftarrow next\_time$
            \STATE $output(\lambda(current\_state))$
            \STATE $current\_state \leftarrow \delta_{int}(current\_state)$
        \ENDIF
        \STATE $last\_time \leftarrow time$
    \ENDWHILE
\end{algorithmic}
\caption{\DEVS simulation pseudo-code for interruptable models.}
\label{algorithm:interruptable}
\end{algorithm*}

\ifnotreport
    \reflection{Why is the output function not called as well before the external transition is invoked?}
\fi

\section{Coupled DEVS models}
\label{sec:coupled}
While our traffic light example is able to receive and output events, there are no other atomic models to communicate with.
To combine different atomic models together and have them communicate, we now introduce coupled models.
This will be done in the context of our previous traffic light, which is connected to a policeman.
The details of the traffic light are exactly like before; the details of the policeman are irrelevant here, as long as it outputs \EAUTO and \EMANUAL events.

\subsection{Basic Coupling}
The first problem we encounter with coupling the traffic light and policeman together is the structure: how do we define a set of models and their interrelations?
This is the core definition of a coupled model: it is merely a structural model that couples models together.
Contrary to the atomic models, there is \textit{no behaviour whatsoever} associated to a coupled model.
Behaviour is the responsibility of atomic models, and structure that of coupled models.

To define the basic structure, we need three elements.
\begin{enumerate}
    \item \textbf{Model instances ($D$)}
        
        The set of model instances defines which models are included within this coupled model.

    \item \textbf{Model specifications ($MS = \{M_i | i \in D\}$)}

        Apart from defining the different instances of submodels, we must include the atomic model specification of these models.
        For each element defined in $D$, we include the 8-tuple specifying the atomic model.
        By definition, a submodel of the coupled \DEVS model always needs to be an atomic model.
        Later on, we will see how this can be extended to support arbitrarily hierarchies.
        \[
            MS = \{M_i | i \in D\} = \{\langle X_i, Y_i, S_i, q_{init, i}, \delta_{int, i}, \delta_{ext, i}, \lambda_i, ta_i \rangle | i \in D\}
        \]
        
    \item \textbf{Model influencees ($IS = \{I_i | i \in D \cup \{\mathit{self}\}\}$)}

        Apart from defining the model instances and their specifications, we need to define the connections between them.
        Connections are defined through the use of influencee sets:
        for each atomic model instance, we define the set of models influenced by that model.
        There are some limitations on couplings, to make sure that inconsistent models cannot be created.
        The following two constraints are imposed:
        \begin{itemize}
            \item \textit{A model should not influence itself.}
                This constraint makes sense, as otherwise it would be possible for a model to influence itself directly.
                While there is no significant problem with this in itself, it would cause the model to trigger both its internal and external transition simultaneously.
                As it is undefined which one should go first, this situation is not allowed.
                In other words, a model should not be an element in its own set of influencees.
                \[
                    \forall i \in D: i \notin I_i
                \]

            \item \textit{Only links within the coupled model are allowed.}
                This is another way of saying that connections should respect modularity.
                Models should not directly influence models outside of the current coupled model, nor models deeper inside of other submodels at this level.
                In other words, the influenced model should be a subset of the set of models in this coupled model.
                \[
                    \forall i \in D: I_i \subseteq D
                \]
        \end{itemize}

        Note that there is no explicit constraint on algebraic loops (\textit{i.e.}, a loop of models that have a time advance equal to zero, preventing the progression of simulated time).
        If this situation is not resolved, it is possible for simulation to get stuck at that specific point in time.
        The situation is only problematic if the circular dependency never gets resolved, causing a livelock of the simulation.
\end{enumerate}
A coupled model can thus be defined as a 3-tuple.
\[
    \langle D, \mathit{MS}, \mathit{IS} \rangle
\]

\subsection{Input and Output}
Our coupled model now couples two atomic models together.
And while it is now possible for the policeman to pass the event to the traffic light, we again lost the ability to send out the state of the traffic light.
The events cannot reach outside of the current coupled model.
Therefore, we need to augment the coupled model with input and output events, which serve as the interface to the coupled model.
This adds the components $X_{\mathit{self}}$ and $Y_{\mathit{self}}$ to the tuple, respectively the set of input and output events, resulting in a 5-tuple.
\[
    \langle X_{\mathit{self}}, Y_{\mathit{self}}, D, MS, IS \rangle
\]

The constraints on the couplings need to be relaxed to accomodate for the new capabilities of the coupled model:
a model can be influenced by the input events of the coupled model, and likewise the models can also influence the output events of the coupled model.
The previously defined constraints are relaxed to allow for $\mathit{self}$, the coupled model itself.
\begin{gather*}
    \forall i \in D \cup \{\mathit{self}\}: i \notin I_i \\
    \forall i \in D \cup \{\mathit{self}\}: I_i \subseteq D \cup \{\mathit{self}\}
\end{gather*}

\subsection{Tie-breaking}
Recall that \DEVS is considered a formal and precise formalism.
But while all components are precisely defined, their interplay is not completely defined yet: what happens when the traffic light changes its state at exactly the same time as the policeman performs its transition?
Would the traffic light switch on to the next state first and then process the policeman's interrupt, or would it directly respond to the interrupt, ignoring the internal event?
While it is a minimal difference in this case, the state reached after the timeout might respond significantly different to the incoming event.

\DEVS solves this problem by defining a \textbf{tie-breaking function ($select$)}.
This function takes all conflicting models and returns the one that gets priority over the others.
After the execution of that internal transition, and possibly the external transitions that it caused elsewhere, it might be that the set of imminent models has changed.
If multiple models are still imminent, we repeat the above procedure (potentially invoking the $select$ function again with the new set of imminent models).
\[
    select : 2^D \rightarrow D
\]

This new addition changes the coupled model to a 6-tuple.
\[
    \langle X_{\mathit{self}}, Y_{\mathit{self}}, D, MS, IS, select \rangle
\]

\subsection{Translation Functions}
Finally, in this case we had full control over both atomic models that are combined.
We might not always be that lucky, as it is possible to reuse atomic models defined elsewhere.
Depending on the application domain of the reused models, they might work with different events.
For example, if our policeman and traffic light were both predefined, with the policeman using \EGOTOWORK and \ETAKEBREAK and the traffic light listening to \EAUTO and \EMANUAL, it would be impossible to directly couple them together.
While it is possible to define wrapper blocks (\textit{i.e.}, artificial atomic models that take an event as input and, with time advance equal to zero, output the translated version), \DEVS provides a more elegant solution to this problem.

Connections are augmented with a \textbf{translation function ($Z_{i,j}$)}, specifying how the event that enters the connection is translated before it is handed over to the endpoint of the connection.
The function thus maps output events to input events, potentially modifying their content.
\begin{center}
\begin{tabular}{lll}
    $Z_{\mathit{self}, j}$ &: $X_{\mathit{self}} \rightarrow X_j$ & $\forall j \in D$ \\
    $Z_{i, \mathit{self}}$ &: $Y_i \rightarrow Y_{\mathit{self}}$ & $\forall i \in D$ \\
    $Z_{i, j}$    &: $Z_i \rightarrow X_j$      & $\forall i, j \in D$ \\
\end{tabular}
\end{center}
These translation functions are defined for each connection, including those between the coupled model's input and output events.
\[
    ZS = \{Z_{i,j} | i \in D \cup \{\mathit{self}\}, j \in I_i\}
\]
The translation function is implicitly assumed to be the identity function if it is not defined.
In case an event needs to traverse multiple connections, all translation functions are chained in order of traversal.

With the addition of this final element, we define a coupled model as a 7-tuple.
\[
    \langle X_{\mathit{self}}, Y_{\mathit{self}}, D, MS, IS, ZS, select \rangle
\]

\subsection{Closure Under Coupling}
Similar to atomic models, we need to formally define the semantics of coupled models.
But instead of explaining the semantics from scratch, by defining some pseudo-code, we map coupled models to equivalent atomic models.
Semantics of a coupled model is thus defined in terms of an atomic model.
In addition, this flattening removes the constraint of coupled models that their submodels should be atomic models: if a coupled model is a submodel, it can be flattened to an atomic model.

In essence, for any coupled model specified as
\[
        <X_{\mathit{self}}, Y_{\mathit{self}}, D, MS, IS, ZS, select>
\]
we define an equivalent atomic model specified as
\[
        <X, Y, S, q_{init}, \delta_{int}, \delta_{ext}, \lambda, ta>
\]
Therefore, we have to define all the elements of the atomic model in terms of elements of the coupled model.
The input and output variables $X$ and $Y$ are easy, since they stay the same.
\begin{gather*}
        X = X_{\mathit{self}} \\
        Y = Y_{\mathit{self}}
\end{gather*}
From an external point of view, the interface of the atomic and coupled model is identical: it has the same input and output events and expects the same kind of data on all of them.

The state $S$ encompasses the parallel composition of the states of all the submodels, including their elapsed times (\textit{i.e.}, the total state $Q$, as defined previously):
\[
        S = \times_{i \in D} Q_i
\]
with the total states $Q_i$ previously defined as:
\[
        Q_i = \{(s_i, e_i)|s_i \in S_i, 0 \leq e_i \leq ta_i(s_i)\}, \forall i \in D
\]
The elapsed time is stored for each model separately, since the elapsed time of the new atomic model updates more frequently than each submodel's elapsed time.

The initial total state is again composed of two components: the initial state ($s_{init}$) and the initial elapsed time ($e_{init}$).
First, we consider the elapsed time ($e_{init}$), which is intuitively equal to the time since the last transition of the flattened model, meaning that it is the minimum of all initial elapsed times, found in all atomic submodels.
Then, the initial state ($s_{init}$) is to be specified, which is the composition of all initial total states of all atomic submodels.
Note, however, that for each initial total state of the submodels, the minimum elapsed time ($e_{init}$) is to be decremented.
Indeed, $q_{init}$ specifies that we entered state $s_{init}$ in the past, more specifically $e_{init}$ time units ago.

\begin{gather*}
        q_{init} = (s_{init}, e_{init}) \\
        s_{init} = (..., (s_{init, i}, e_{init, i} - e_{init}), ...) \\
        e_{init} = min_{i \in D} \{e_{init, i}\} 
\end{gather*}

The time advance function $ta$ then returns the minimum of all remaining times.
\[
        ta(s) = min \{\sigma_i = ta_i(s_i) - e_i|i \in D\}
\]
The imminent component is chosen from the set of all models with the specified minimum remaining time ($IMM$).
This set contains all models whose remaining time ($\sigma_i$) is identical to the time advance of the flattened model ($ta$).
The $select$ function is then used to reduce this set to a single element $i^*$.
\begin{gather*}
        IMM(s) = \{i \in D|\sigma_i = ta(s)\} \\
        i^* = \mathit{select}(IMM(s))
\end{gather*}

The output function $\lambda$ executes the output function of $i^*$ and applies the translation function, but only if the model influences the flattened model directly (\textit{i.e.}, if the output of $i^*$ is routed to the coupled model's output).
If there is no connection to the coupled model's output (\textit{i.e.}, $i^*$ is only coupled to other atomic models), no output function is invoked here.
We will see later on that these events are still generated, but they are consumed internally elsewhere.
\[
        \lambda(s) = \left\{ \begin{array}{l l}
                Z_{i^*, \mathit{self}}(\lambda_{i^*}(s_{i^*})) & \quad \text{if } \mathit{self} \in I_{i^*} \\
                \phi & \quad \text{otherwise} \\
                \end{array} \right.
\]

The \emph{internal transition function} is defined for each part of the state separately:
\[
        \delta_{int}(s) = (\dots, (s'_j, e'_j), \dots)
\]
With three kinds of models:
(1) the model $i^*$ itself, which just performs its internal transition function;
(2) the models influenced by $i^*$, which perform their external transition based on the output generated by $i^*$;
(3) models unrelated to $i^*$.
In all cases, the elapsed time is updated.
\[
        (s'_j, e'_j) = \left\{\begin{array}{l l}
                (\delta_{int, j}(s_j), 0) & \quad \text{for } j = i^*, \\
                (\delta_{ext, j}((s_j, e_j + ta(s)), Z_{i^*, j}(\lambda_{i^*}(s_{i^*}))), 0) & \quad \text{for } j \in I_{i^*}, \\
                (s_j, e_j + ta(s)) & \quad \text{otherwise} \\
        \end{array} \right.
\]
Note that the internal transition function includes external transition functions of submodels for those models influenced by $i^*$.
As $i^*$ outputs events that are consumed internally, this all happens internally.

The \emph{external transition} function is similar to the internal transition function.
Now two types are distinguished:
(1) models directly connected to the input of the model, which perform their external transition;
(2) models not directly connected to the input of the model, which only update their elapsed time.
\[
        \delta_{ext}((s, e), x) = (\dots, (s'_i, e'_i), \dots)
\]
\[
        (s'_i, e'_i) = \left\{\begin{array}{l l}
                (\delta_{ext, i}((s_i, e_i + e), Z_{\mathit{self}, i}(x)), 0) & \quad \text{for } i \in I_{\mathit{self}} \\
                (s_i, e_i + e) & \quad \text{otherwise}
        \end{array} \right.
\]

\ifnotreport
    \reflection{Is it possible to replace the translation function $Z$ by an atomic DEVS model that takes input and puts the translated value on its output, while preserving semantics?}
\fi

\section{The DEVS Abstract Simulator}
\label{sec:semantics}
Up to now, the semantics of atomic models was defined through natural language and high-level pseudo-code.
Coupled models were given semantics through a mapping to these atomic models.
Both of these have their own problems.
For atomic models, the pseudo-code is not sufficiently specific to create a compliant \DEVS simulator: a lot of details of the algorithm are left unspecified (\textit{e.g.}, where does the external event come from).
For coupled models, the flattening procedure is elegant and formal, though it is highly inefficient to perform this flattening at run-time.

To counter these problems, we will define a more elaborate, and formal, simulation algorithm for both atomic and coupled models.
Atomic models get a more specific definition with a clear interface, and coupled models get their own simulation algorithm without flattening.
Coupled models are thus given ``operational semantics'' instead of ``translational semantics''.

This simulation algorithm, an \textit{abstract simulator} forms the basis for more efficient simulation algorithms, and serves as a reference algorithm.
Its goal is to formally define the semantics of both models in a concise way, without caring about performance or implementation issues.
Adaptations are allowed, but the final result should be identical: simulation results are to be completely independent from the implementation.
A direct implementation of the abstract simulator is inefficient, and actual implementations therefore vary significantly.

We now elaborate on the abstract simulator algorithm.
For each atomic and coupled model, an instance is created of the respective algorithm.

Table~\ref{table:variables} shows the different variables used, their type, and a brief explanation.

\begin{longtable}{l c l}
    \caption{Variables used in the abstract simulator.}         \\
    \label{table:variables}
        \textbf{name} & \textbf{type} & \textbf{explanation}    \\
        \hline
        \hline
        $t_l$   & time  & simulation time of last transition    \\
        $t_n$   & time  & simulation time of next transition    \\
        $t$     & time  & current simulation time               \\
        $e$     & time  & elapsed time since last transition    \\
        $s$     & state & current state of the atomic model     \\
        $x$     & event & incoming event                        \\
        $y$     & event & outgoing event                        \\
        $from$  & model & source of the incoming message        \\
        $parent$& model & coupled model containing this model   \\
        $self$  & model & current model                         \\
\end{longtable}

We furthermore distinguish five types of synchronization messages, as exchanged between the different abstract simulators.
An overview of messages is shown in Table~\ref{table:synchronization}.
\begin{longtable}{l l}
    \caption{Types of synchronization messages.}            \\
    \label{table:synchronization}
        \textbf{type}   & \textbf{explanation}              \\
        \hline
        \hline
        $i$             & initialization of the simulation  \\
        $*$             & transition in the model           \\
        $x$             & input event for the model         \\
        $y$             & output event from the model       \\
        $done$          & computation finished for a model  \\
\end{longtable}

First is the abstract simulation algorithm for atomic models, presented in Algorithm~\ref{algorithm:classic_atomic}.
This algorithm consists of a big conditional, depending on the message that is received.
Atomic models only perform an operation upon reception of a message: there is no autonomous behaviour.
This algorithm is invoked every time a synchronization message is received.
Messages consist of three components: the type of the message, the source of the message, and the simulation time.
The conditional consists of three options:
On the reception of an $i$ message, we perform \textit{initialization} of the simulation time.

Another option is the reception of a $*$ message, triggering a \textit{transition}.
The message consists of both a sender and the time at which the transition should happen.
By definition, a transition can only happen at time $t_n$, so we assert this.
After this check, we have to perform the following steps:
(1) generate the output, (2) send it out to the sender of the $*$ message (our parent), (3) perform the internal transition, (4) update our time with the time advance, and (5) indicate to our parent that we finished processing the message, also passing along our time of next transition.

Finally, it is possible to receive an $x$ message, indicating \textit{external input}.
This can happen anytime between our last transition ($t_l$), and our scheduled transition ($t_n$), so we again assert the simulation time.
Note that these times are inclusive: due to the $select$ function it is possible that another model comes right after or before our own scheduled transition.
We perform the following steps:
(1) compute the elapsed time ($e$) based on the provided simulation time ($t$), (2) perform the external transition, (3) update the simulation time of the next transition, and (4) indicate to our parent that we finished processing the message, also passing along our time of next transition.

\begin{algorithm*}
\begin{algorithmic}
    \IF {receive $(i, from, t)$ message}
        \STATE $t_l \leftarrow t - e$
        \STATE $t_n \leftarrow t_l + ta(s)$
        \STATE send $(done, self, t_n)$ to $parent$
    \ELSIF {receive $(*, from, t)$ message}
        \IF {$t = t_n$}
            \STATE $y \leftarrow \lambda(s)$
            \IF {$ y \not= \phi$}
                \STATE send $(y, self, t)$ to $parent$
            \ENDIF
            \STATE $s \leftarrow \delta_{int}(s)$
            \STATE $t_l \leftarrow t$
            \STATE $t_n \leftarrow t_l + ta(s)$
            \STATE send $(done, self, t_n)$ to parent
        \ENDIF
    \ELSIF {receive $(x, from, t)$ message}
        \IF {$t_l \leq t \leq t_n$}
            \STATE $e \leftarrow t - t_l$
            \STATE $s \leftarrow \delta_{ext}((s, e), x)$
            \STATE $t_l \leftarrow t$
            \STATE $t_n \leftarrow t_l + ta(s)$
            \STATE send $(done, self, t_n)$ to $parent$
        \ELSE
            \STATE error: bad synchronization
        \ENDIF
    \ENDIF
\end{algorithmic}
\caption{\DEVS atomic model abstract simulator.}
\label{algorithm:classic_atomic}
\end{algorithm*}

Recall that the abstract simulation algorithm did not have any autonomous behaviour.
This indicates that there is another entity governing the progression of the simulation
This simulation entity is the root coordinator, and it encodes the main simulation loop.
Its algorithm is shown in Algorithm~\ref{algorithm:classic_root}.
As long as simulation needs to continue, it sends out a message to the topmost model in the hierarchy to perform transitions.
When a reply is received, simulation time is progressed to the time indicated by the topmost model.

\begin{algorithm*}
\begin{algorithmic}
    \STATE send $(i, main, 0.0)$ to topmost coupled model $top$
    \STATE wait for $(done, top, t_N)$
    \STATE $t \leftarrow t_N$
    \WHILE{\NOT $terminationCondition()$}
        \STATE send $(*, main, t)$ to topmost coupled model $top$
        \STATE wait for $(done, top, t_N)$
        \STATE $t \leftarrow t_N$
    \ENDWHILE
\end{algorithmic}
\caption{\DEVS root coordinator.}
\label{algorithm:classic_root}
\end{algorithm*}

Finally, while not completely necessary due to the existence of the flattening algorithm, we also define a shortcut for the simulation of coupled models.
The abstract simulation algorithm for coupled models is shown in Algorithm~\ref{algorithm:classic_coupled}.
Coupled models can receive all five different types of synchronization messages.

First, the $i$ message again indicates \textit{initialization}.
It merely forwards the message to all of its children and marks each child as active.
Every coupled model has a $t_l$ and $t_n$ variable as well, which is defined as the maximum, respectively minimum, of its children.
This is logical, as any transition of its children will also require an operation on the coupled model containing it.
When a message is sent to a submodel, the submodel is marked as active.
The use for this is shown in the processing of the $done$ message.

Second, the $*$ message again indicates a \textit{transition}.
Contrary to the atomic models, a coupled model is unable to perform a transition itself.
Instead, it forwards the message to the imminent submodel, found by executing the $select$ function for all models that have that exact same $t_n$.
Only a single model will be selected, and a $*$ message is sent to that model.
Just like before, the model is marked as active to make sure that we wait for its computation to finish.

Third, a $y$ message indicates an \textit{output} message.
The output message is output by the output function of a subcomponent, and needs to be routed through the coupled model.
This part of the function is responsible for routing the message to the influencees of the model that sent out the message.
Note that it is also possible that one of the influencees is $self$, indicating that the message needs to be routed externally (\textit{i.e.}, to the output of the coupled model).
In any case, the message needs to be translated using the translation function.
The actual translation function that is invoked depends on the source and destination of the message.

Fourth, a $x$ message can be received, indicating \textit{input}.
This is mostly identical to the output messages, only now can we also handle messages that were received from our own parent.

Finally, a $done$ message can be received, indicating that a submodel has \textit{finished} its computation.
The submodel, which was marked as an active child, will now be unmarked.
When $done$ messages are received from all submodels (\textit{i.e.}, all children are inactive), we determine our own $t_l$ and $t_n$ variables and send out the minimal $t_n$ of all submodels.
This time is then sent to the parent.

The abstract simulator for coupled models can work with any kind of submodel, not necessarily atomic models.
In deep hierarchies, the $done$ message always propagates the minimal $t_n$ upwards in the hierarchy.
In the end, the root coordinator will always receive the minimal $t_n$, which is the time of the earliest next internal transition.

\begin{algorithm*}
\begin{algorithmic}
    \IF {receive $(i, from, t)$ message}
        \FORALL{$d$ in $D$}
            \STATE send $(i, self, t)$ to $d$
            \STATE $active\_children \leftarrow active\_children \cup \{d\}$
        \ENDFOR
    \ELSIF {receive $(*, from, t)$ message}
        \IF {$t = t_n$}
            \STATE $i* = select(\{M_i.t_n = t | i \in D\})$
            \STATE send $(*, self, t)$ to $i*$
            \STATE $active\_children \leftarrow active\_children \cup \{i^*\}$
        \ENDIF
    \ELSIF {receive $(y, from, t)$ message}
        \FORALL {$i \in I_{from} \setminus \{self\}$}
            \STATE send $(Z_{from, i}(y), from, to)$ to $i$
            \STATE $active\_children \leftarrow active\_children \cup \{i\}$
        \ENDFOR
        \IF {$self \in I_{from}$}
            \STATE send $(Z_{from, self}(y), self, t)$ to $parent$
        \ENDIF
    \ELSIF {receive $(x, from, t)$ message}
        \IF {$t_l \leq t \leq t_n$}
            \FORALL {$i \in I_{from}$}
                \STATE send $(Z_{self, i}(x), self, t)$ to $i$
                \STATE $active\_children \leftarrow active\_children \cup \{i\}$
            \ENDFOR
        \ENDIF
    \ELSIF {receive $(done, from, t)$ message}
        \STATE $active\_children \leftarrow active\_children \setminus \{from\}$
        \IF {$active\_children = \phi$}
            \STATE $t_l \leftarrow max\{t_{l,d} | d \in D\}$
            \STATE $t_n \leftarrow min\{t_{n,d} | d \in D\}$
            \STATE send $(done, self, t_n)$ to $parent$
        \ENDIF
    \ENDIF
\end{algorithmic}
\caption{\DEVS coupled model abstract simulator.}
\label{algorithm:classic_coupled}
\end{algorithm*}

\ifnotreport
    \reflection{Is it possible for an atomic DEVS model to do an internal and external transition at the same point in simulated time? Explain your answer.}
\fi

\section{Application to Queueing Systems}
\label{sec:applications}
The usefulness of \DEVS of course goes further than traffic lights.
To present a more realistic model and highlight the potential for performance analysis, we present a simple queueing system next.
While a lot has been done in queueing theory, we present simulation as an alternative to the mathematical solutions.
Even though the mathematical solutions have their advantages, simulation offers more flexibility and does not get that complex.
It is, however, necessarily limited to ``sampling'': simulations will only take samples and will therefore generally not find rare and exceptional cases.
Not taking them into account is fine in many situations, as it is now in our example model.

In this section, we present a simple queueing problem.
Variations on this model --- in either its behaviour, structure, or parameters --- are easy to do.

\subsection{Problem Description}
In this example, we model the behaviour of a simple queue that gets served by multiple processors.
Implementations of this queueing systems are widespread, such as for example at airport security.
Our model is parameterizable in several ways: we can define the random distribution used for event generation times and event size, the number of processors, performance of each individual processor, and the scheduling policy of the queue when selecting a processor.
Clearly, it is easier to implement this, and all its variants, in \DEVS than it is to model it mathematically.
For our performance analysis, we show the influence of the number of processors (\textit{e.g.}, metal detectors) on the average and maximal queueing time of jobs (\textit{e.g.}, travellers).

\begin{figure}
    \center
    \includegraphics[width=0.6\textwidth]{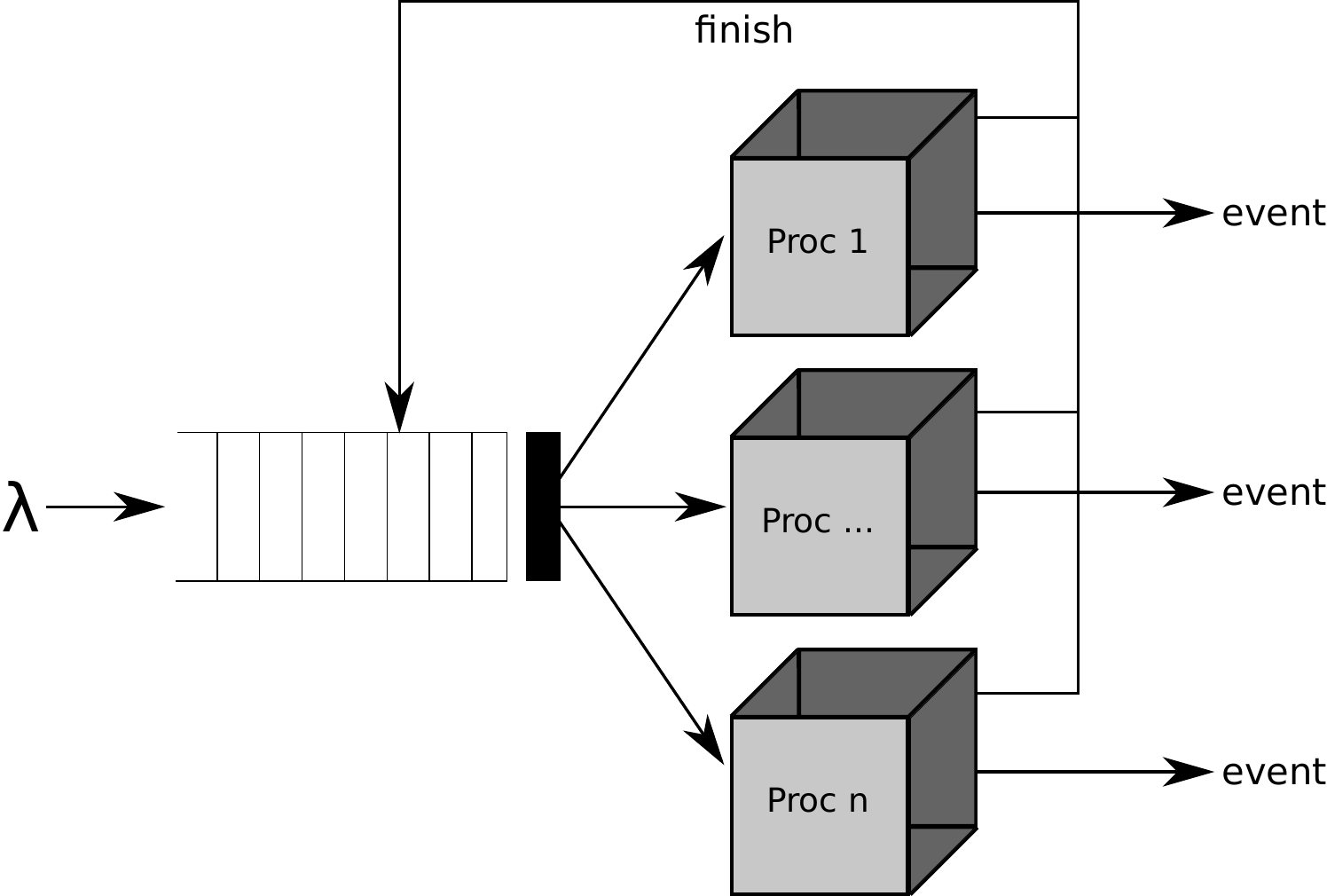}
    \caption{Queue system with a single generator, single queue, and $n$ processors.}
    \label{fig:queue}
\end{figure}

A model of this system can be shown in Figure~\ref{fig:queue}.
Events (people) are generated by a generator using some distribution function.
They enter the queue, which decides the processor that they will be sent to.
If multiple processors are available, it picks the processor that has been idle for the longest; if no processors are available, the event is queued until a processor becomes available.
The queue works First-In-First-Out (FIFO) in case multiple events are queueing.
For a processor to signal that it is available, it needs to signal the queue.
The queue keeps track of available processors.
When an event arrives at a processor, it is processed for some time, depending on the size of the event and the performance characteristics of the processor.
After processing, the processor signals the queue and sends out the event that was being processed.

\subsection{Description in \DEVS}
While examples could be given purely in their formal description, they would not be executable and would introduce a significant amount of accidental complexity.
We use the tool PythonPDEVS\footnote{Download: \url{http://msdl.cs.mcgill.ca/projects/DEVS/PythonPDEVS}}~\cite{PythonPDEVS1,JDF} to implement the \DEVS model and perform simulations.
In PythonPDEVS, \DEVS models are implemented by defining methods that implement the different aspects of the tuple.
All code within these methods is just normal Python code, though a minimal number of API calls is required in the case of a coupled \DEVS model.
Since most \DEVS tools work similarly, these examples could easily be transposed to other \DEVS simulation tools.
An overview of popular DEVS simulation tools is shown in~\cite{DEVSSurvey}.

To specify this model, we first define the event exchanged between different models: the \textit{Job}.
A job is coded as a class \texttt{Job}.
It has the attributes \textit{size} (\textit{i.e.}, indicative of processing time) and \textit{creation time} (\textit{i.e.}, for statistic gathering).
The \texttt{Job} class definition is shown in Listing~\ref{lst:job}.

\begin{lstlisting}[label=lst:job,caption={PythonPDEVS code for the \texttt{Job} event.}]
class Job:
    def __init__(self, size, creation_time):
        # Jobs have a size and creation_time parameter
        self.size = size
        self.creation_time = creation_time
\end{lstlisting}

We now focus on each atomic model separately, starting at the event generator.

The \textit{generator} is defined as an atomic model using the class \texttt{Generator}, shown in Listing~\ref{lst:generator}.
Classes that represent an atomic model inherit from the \texttt{AtomicDEVS} class.
They should implement methods that implement each of the \DEVS components.
Default implementations are provided for a passivated model, such that unused functions do not need to be defined.
In the constructor, input and output ports are defined, as well as model parameters and the initial state.
We see that the definition of the generator is very simple: we compute the time remaining until the next event (\texttt{remaining}), and decrement the number of events to send.
The generator also keeps track of the current simulation time, in order to set the creation time of events.
The time advance function returns the time remaining until the next internal transition.
Finally, the output function returns a new customer event with a randomly defined size.
The job has an attribute containing the time at which it was generated.
Recall, however, that the output function was invoked before the internal transition, so the current time has not yet been updated by the internal transition.
Therefore, the output function also has to do this addition, without storing the result in the state (as it cannot write to the state).

\begin{lstlisting}[label=lst:generator,caption={PythonPDEVS code for the Generator atomic model.}]
from pypdevs.DEVS import AtomicDEVS
from job import Job
import random

# Define the state of the generator as a structured object
class GeneratorState:
    def __init__(self, gen_num):
        # Current simulation time (statistics)
        self.current_time = 0.0
        # Remaining time until generation of new event
        self.remaining = 0.0
        # Counter on how many events to generate still
        self.to_generate = gen_num

class Generator(AtomicDEVS):
    def __init__(self, gen_param, size_param, gen_num):
        AtomicDEVS.__init__(self, "Generator")
        # Output port for the event
        self.out_event = self.addOutPort("out_event")
        # Define the state
        self.state = GeneratorState(gen_num)

        # Parameters defining the generator's behaviour
        self.gen_param = gen_param
        self.size_param = size_param

    def intTransition(self):
        # Update simulation time
        self.state.current_time += self.timeAdvance()
        # Update number of generated events
        self.state.to_generate -= 1
        if self.state.to_generate == 0:
            # Already generated enough events, so stop
            self.state.remaining = float('inf')
        else:
            # Still have to generate events, so sample for new duration
            self.state.remaining = random.expovariate(self.gen_param)
        return self.state

    def timeAdvance(self):
        # Return remaining time; infinity when generated enough
        return self.state.remaining

    def outputFnc(self):
        # Determine size of the event to generate
        size = max(1, int(random.gauss(self.size_param, 5)))
        # Calculate current time (note the addition!)
        creation = self.state.current_time + self.state.remaining
        # Output the new event on the output port
        return {self.out_event: Job(size, creation)}
\end{lstlisting}

Next up is the queue, which is the most interesting component of the simulation, as it is the part we wish to analyze.
The \texttt{Queue} implementation is similar in structure to the \texttt{Generator}.
Of course, the \DEVS parts get a different specification, as shown in Listing~\ref{lst:queue}.
The queue takes a structural parameter, specifying the number of processors.
This is needed since the queue has an output port for each processor.
When an internal transition happens, the queue knows that it has just output an event to the first idle processor.
It thus marks the first idle processor as busy, and removes the event it was currently processing.
If there are events remaining in the queue, and a processor is available to process it, we process the first element from the queue and set the \texttt{remaining\_time} counter.
In the external transition, we check the port we received the event on.
Either it is a signal of the processor to indicate that it has finished, or else it is a new event to queue.
In the former case, we mark the processor that sent the event as idle, and potentially process a queued message.
For this to work, the processor should include its ID in the event, as otherwise the queue has no idea who sent this message.
In the latter case, we either process the event immediately if there are idle processors, or we store it in the queue.
The time advance merely has to return the \texttt{remaining\_time} counter that is managed in both transition functions.
Finally in the output function, the model outputs the first queued event to the first available processor.
Note that we can only read the events and processors, and cannot modify these lists: state modification is reserved for the transition functions.
An important consideration in this model is the \texttt{remaining\_time} counter, which indicates how much time remains before the event is processed.
We cannot simply put the processing time of events in the time advance, as interrupts could happen during this time.
When an interrupt happens (\textit{e.g.}, another event arrives), the time advance is invoked again, and would return the total processing time, instead of the remaining time to process the event.
To solve this problem, we maintain a counter that explicitly gets decremented when an external interrupt happens.

\begin{lstlisting}[label=lst:queue,caption={PythonPDEVS code for the Queue atomic model.}]
from pypdevs.DEVS import AtomicDEVS

# Define the state of the queue as a structured object
class QueueState:
    def __init__(self, outputs):
        # Keep a list of all idle processors
        self.idle_procs = range(outputs)
        # Keep a list that is the actual queue data structure
        self.queue = []
        # Keep the process that is currently being processed
        self.processing = None
        # Time remaining for this event
        self.remaining_time = float("inf")

class Queue(AtomicDEVS):
    def __init__(self, outputs):
        AtomicDEVS.__init__(self, "Queue")
        # Fix the time needed to process a single event
        self.processing_time = 1.0
        self.state = QueueState(outputs)

        # Create 'outputs' output ports
        # 'outputs' is a structural parameter!
        self.out_proc = []
        for i in range(outputs):
            self.out_proc.append(self.addOutPort("proc_%i" % i))

        # Add the other ports: incoming events and finished event
        self.in_event = self.addInPort("in_event")
        self.in_finish = self.addInPort("in_finish")

    def intTransition(self):
        # Is only called when we are outputting an event
        # Pop the first idle processor and clear processing event
        self.state.idle_procs.pop(0)
        if self.state.queue and self.state.idle_procs:
            # There are still queued elements, so continue
            self.state.processing = self.state.queue.pop(0)
            self.state.remaining_time = self.processing_time
        else:
            # No events left to process, so become idle
            self.state.processing = None
            self.state.remaining_time = float("inf")
        return self.state

    def extTransition(self, inputs):
        # Update the remaining time of this job
        self.state.remaining_time -= self.elapsed
        # Several possibilities
        if self.in_finish in inputs:
            # Processing a "finished" event, so mark proc as idle
            self.state.idle_procs.append(inputs[self.in_finish])
            if not self.state.processing and self.state.queue:
                # Process first task in queue
                self.state.processing = self.state.queue.pop(0)
                self.state.remaining_time = self.processing_time
        elif self.in_event in inputs:
            # Processing an incoming event
            if self.state.idle_procs and not self.state.processing:
                # Process when idle processors
                self.state.processing = inputs[self.in_event]
                self.state.remaining_time = self.processing_time
            else:
                # No idle processors, so queue it
                self.state.queue.append(inputs[self.in_event])
        return self.state

    def timeAdvance(self):
        # Just return the remaining time for this event (or infinity else)
        return self.state.remaining_time

    def outputFnc(self):
        # Output the event to the processor
        port = self.out_proc[self.state.idle_procs[0]]
        return {port: self.state.processing}
\end{lstlisting}

The next atomic model is the \texttt{Processor} class, shown in Listing~\ref{lst:processor}.
It merely receives an incoming event and starts processing it.
Processing time, computed upon receiving an event in the external transition, is dependent on the size of the task, but takes into account the processing speed and a minimum amount of processing that needs to be done.
After the task is processed, we trigger our output function and internal transition function.
We need to send out two events: one containing the job that was processed, and one to signal the queue that we have become available.
For this, two different ports are used.
Note that the definition of the processor would not be this simple in case there was no queue before it.
We can now make the assumption that when we get an event, we are already idle and therefore don't need to queue new incoming events first.

\begin{lstlisting}[label=lst:processor,caption={PythonPDEVS code for the Processor atomic model.}]
from pypdevs.DEVS import AtomicDEVS

# Define the state of the processor as a structured object
class ProcessorState(object):
    def __init__(self):
        # State only contains the current event
        self.evt = None

class Processor(AtomicDEVS):
    def __init__(self, nr, proc_param):
        AtomicDEVS.__init__(self, "Processor_%i" % nr)

        self.state = ProcessorState()
        self.in_proc = self.addInPort("in_proc")
        self.out_proc = self.addOutPort("out_proc")
        self.out_finished = self.addOutPort("out_finished")

        # Define the parameters of the model
        self.speed = proc_param
        self.nr = nr

    def intTransition(self):
        # Just clear processing event
        self.state.evt = None
        return self.state

    def extTransition(self, inputs):
        # Received a new event, so start processing it
        self.state.evt = inputs[self.in_proc]
        # Calculate how long it will be processed
        time = 20.0 + max(1.0, self.state.evt.size / self.speed)
        self.state.evt.processing_time = time
        return self.state

    def timeAdvance(self):
        if self.state.evt:
            # Currently processing, so wait for that
            return self.state.evt.processing_time
        else:
            # Idle, so don't do anything
            return float('inf')

    def outputFnc(self):
        # Output the processed event and signal as finished
        return {self.out_proc: self.state.evt,
                self.out_finished: self.nr}
\end{lstlisting}

The processor finally sends the task to the \texttt{Collector} class, shown in Listing~\ref{lst:collector}.
The collector is an artificial component that is not present in the system being modeled; it is only used for statistics gathering.
For each job, it stores the time in the queue.

\begin{lstlisting}[label=lst:collector,caption={PythonPDEVS code for the Collector atomic model.}]
from pypdevs.DEVS import AtomicDEVS

# Define the state of the collector as a structured object
class CollectorState(object):
    def __init__(self):
        # Contains received events and simulation time
        self.events = []
        self.current_time = 0.0

class Collector(AtomicDEVS):
    def __init__(self):
        AtomicDEVS.__init__(self, "Collector")
        self.state = CollectorState()
        # Has only one input port
        self.in_event = self.addInPort("in_event")

    def extTransition(self, inputs):
        # Update simulation time
        self.state.current_time += self.elapsed
        # Calculate time in queue
        evt = inputs[self.in_event]
        time = self.state.current_time - evt.creation_time - evt.processing_time
        inputs[self.in_event].queueing_time = max(0.0, time)
        # Add incoming event to received events
        self.state.events.append(inputs[self.in_event])
        return self.state

    # Don't define anything else, as we only store events.
    # Collector has no behaviour of its own.
\end{lstlisting}

With all atomic models defined, we only have to couple them together in a coupled model, as shown in Listing~\ref{lst:system}.
In this system, we instantiate a generator, queue, and collector, as well as a variable number of processors.
The number of processors is variable, but is still static during simulation.
The couplings also depend on the number of processors, as each processor is connected to the queue and the collector.

\begin{lstlisting}[label=lst:system,caption={PythonPDEVS code for the System coupled model.}]
from pypdevs.DEVS import CoupledDEVS

# Import all models to couple
from generator import Generator
from queue import Queue
from processor import Processor
from collector import Collector

class QueueSystem(CoupledDEVS):
    def __init__(self, mu, size, num, procs):
        CoupledDEVS.__init__(self, "QueueSystem")

        # Define all atomic submodels of which there are only one
        generator = self.addSubModel(Generator(mu, size, num))
        queue = self.addSubModel(Queue(len(procs)))
        collector = self.addSubModel(Collector())

        self.connectPorts(generator.out_event, queue.in_event)

        # Instantiate desired number of processors and connect
        processors = []
        for i, param in enumerate(procs):
            processors.append(self.addSubModel(
                              Processor(i, param)))
            self.connectPorts(queue.out_proc[i],
                              processors[i].in_proc)
            self.connectPorts(processors[i].out_finished,
                              queue.in_finish)
            self.connectPorts(processors[i].out_proc,
                              collector.in_event)

        # Make it accessible outside of our own scope
        self.collector = collector
\end{lstlisting}

Now that our \DEVS model is completely specified, we can start running simulations on it.
Simulation requires an \textit{experiment} file though, which initializes the model with parameters and defines the simulation configuration.
An example experiment, again in Python, is shown in Listing~\ref{lst:experiment}.
The experiment writes out the raw queueing times to a Comma Seperated Value (CSV) file.
An experiment file often contains some configuration of the simulation tool, which differs for each tool.
For PythonPDEVS, the documentation\footnote{\url{http://msdl.cs.mcgill.ca/projects/DEVS/PythonPDEVS/documentation/html/index.html}} provides an overview of supported options.

\begin{lstlisting}[label=lst:experiment,caption={PythonPDEVS code for the experiment on the system.}]
from pypdevs.simulator import Simulator
import random

# Import the model we experiment with
from system import QueueSystem

# Configuration:
# 1) number of customers to simulate
num = 500
# 2) average time between two customers
time = 30.0
# 3) average size of customer
size = 20.0
# 4) efficiency of processors (products/second)
speed = 0.5
# 5) maximum number of processors used
max_processors = 10
# End of configuration

# Store all results for output to file
values = []
# Loop over different configurations
for i in range(1, max_processors):
    # Make sure each of them simulates exactly the same workload
    random.seed(1)
    # Set up the system
    procs = [speed] * i
    m = QueueSystem(mu=1.0/time, size=size, num=num, procs=procs)

    # PythonPDEVS specific setup and configuration
    sim = Simulator(m)
    sim.setClassicDEVS()
    sim.simulate()

    # Gather information for output
    evt_list = m.collector.state.events
    values.append([e.queueing_time for e in evt_list])

# Write data to file
with open('output.csv', 'w') as f:
    for i in range(num):
        f.write("%s" % i)
        for j in range(len(values)):
            f.write(", %5f" % (values[j][i]))
        f.write("\n")
\end{lstlisting}

\subsection{Performance Analysis}
After the definition of our \DEVS model and experiment, we of course still need to run the simulation.
Simply by executing the experiment file, the CSV file is generated, and can be analyzed in a spreadsheet tool or plotting library.
Depending on the data stored during simulation, analysis can show the average queueing times, maximal queueing times, number of events, processor utilization, and so on.

\begin{figure}
    \center
    \begin{minipage}{0.48\textwidth}
        \center
        \includegraphics[width=\textwidth]{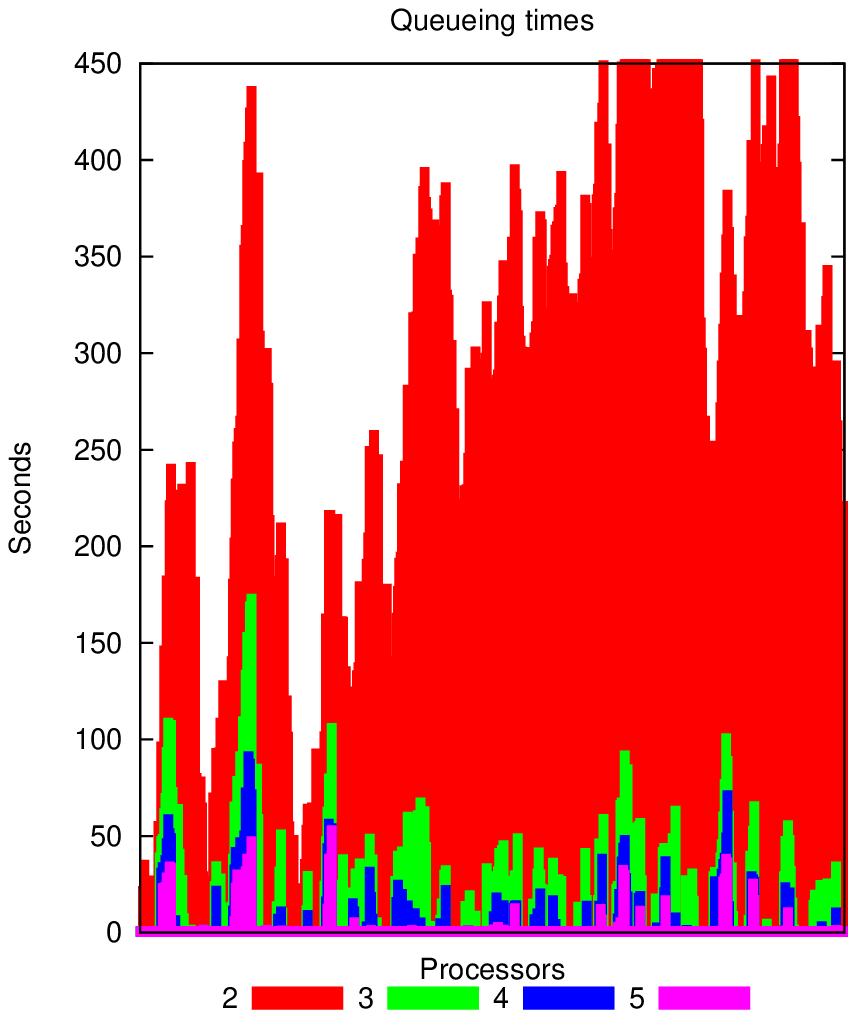}
        \caption{Evolution of queueing times for subsequent customers.}
        \label{fig:queueing_evolution}
    \end{minipage}
    \begin{minipage}{0.48\textwidth}
        \center
        \includegraphics[width=\textwidth]{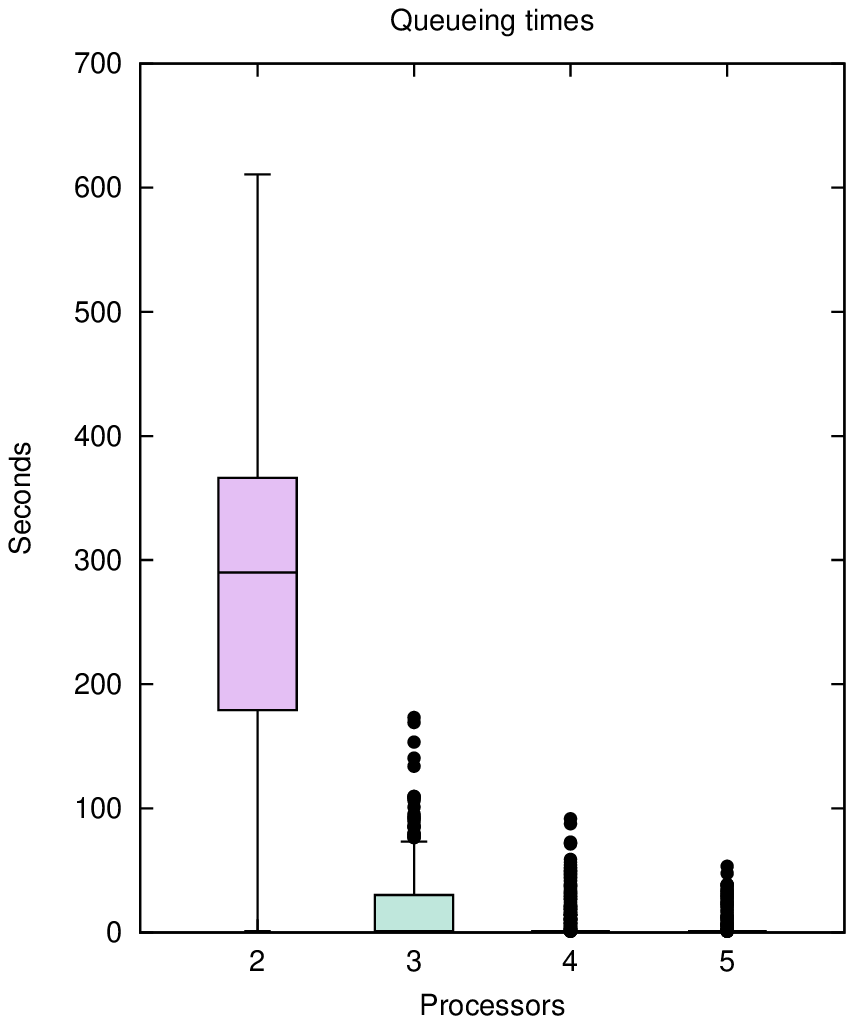}
        \caption{Boxplot of queueing times for varying number of active processors.}
        \label{fig:queueing_boxplot}
    \end{minipage}
\end{figure}

Corresponding to our initial goal, we perform the simulation in order to find out the influence of opening multiple processors on the average and maximum queueing time.
Figure~\ref{fig:queueing_evolution} shows the evolution of the waiting time for subsequent clients.
Figure~\ref{fig:queueing_boxplot} shows the same results, drawn using boxplots.
These results indicate that while two processors are able to handle the load, maximum waiting time is rather high: a median of 200 seconds and a maximum of around 470 seconds.
When a single additional processor is added, average waiting time decreases significantly, and the maximum waiting time also becomes tolerable: the mean job is served immediately, with 75\% of jobs being handled within 25 seconds.
Further adding processors still has a positive effect on queueing times, but the effect might not warrant the increased cost in opening processors: apart from some exceptions, all customers are processed immediately starting from four processors.
Ideally, a cost function would be defined to quantize the value (or dissatisfaction) of waiting jobs, and compare this to the cost of adding additional processors.
We can then optimize that cost function to find out the ideal balance between paying more for additional processors and losing money due to long job processing times.
Of course, this ideal balance depends on several factors, including our model configuration and the cost function used.

\ifnotreport
    \reflection{What would you have to change in order to use a different queueing discipline or arrival process? How does this compare to the changes needed when using a mathematical model of this same process?}
\fi

\section{Variants}
\label{sec:variants}
Despite the success of the original \DEVS specification, as introduced throughout this chapter, shortcoming were identified when used in some domains.
For these reasons, a lot of variants have recently spawned.
In this section, we touch upon the three most popular ones, with some remarks on other variants.
Note that we make the distinction between variants that further augment the \DEVS formalism (\textit{i.e.}, make more constructs available), and those that restrict it (\textit{i.e.}, prevent several cases).
Both have their reasons, mostly related to the implementation:
augmenting the \DEVS formalism makes it easier for modellers to create models in some domains,
whereas limiting the \DEVS formalism makes some operations, such as analysis, possible or easier to implement.

\subsection{Parallel DEVS}
One of the main problems identified in \DEVS is related to performance: when multiple models are imminent, they are processed sequentially.
While \DEVS does allow for some parallelism, (\textit{e.g.}, between simultaneous external transitions), multiple internal transitions is a common occurence.

\PDEVS~\cite{ParallelDEVS} was introduced as a variant of the \DEVS formalism, in which parallel execution of internal transition functions is allowed.
This changes the semantics of models though, so it requires changes to the abstract simulator~\cite{ParallelDEVSAbsSim}.
The proposed changes are therefore not just syntactic sugar: they explicitly modify the semantics of models.

Allowing for parallelism is, however, not a trivial task: several modifications are required, which we briefly mention here.
The first logical change is the removal of the $select$ function: instead of selecting a model to go first, all imminent models are allowed to transition simultaneously.
Whether or not this happens in parallel or not, as it might not necessarily be faster~\cite{SequentialPDEVS}, is up to the implementation.
This creates some repercussions throughout the remainder of the formalism, as the $select$ function was introduced for good reasons.

Since models can now perform their internal transition simultaneously, output functions also happen simultaneously.
While this is in itself not a problem, routing might cause the need for events to be merged together, for example when two events get routed to the same model.
The abstract simulator was not designed for this, as an external transition was immediately invoked upon the reception of an external event.
So in \PDEVS, events are always encapsulated in \textit{bags}, which can easily be merged.
Bags, or multisets, are a kind of set which can contain items multiple times.
This way, multiple bags can be trivially joined, without losing information.
Note that order is undefined, as otherwise it would depend on the synchronization between different output functions: which one is executed before the other.
Due to this change in interface, the external transition needs to be altered to operate on a bag of input events, and the output function has to generate a bag of output events.

Problems don't stop there, as internal and external transition might happen simultaneously.
Recall that in \DEVS, self-loops were not allowed for this exact purpose.
In \PDEVS, however, two models can perform their internal transition simultaneously, with one outputting an event for the other one.
In that case, the model needs to process both its internal transition, and its external transition caused by the other model's transition.
Since there is no priority defined between them (that was part of the purpose of the $select$ function), they should execute simultaneously.
To allow for this, a new kind of transition is defined: the \textit{confluent transition function}.
This transition is only performed when both the internal and external transition happen simultaneously.
\PDEVS leaves open what the semantics of this is, though a sane default is for the internal transition to go first, followed by the external transition.

Thanks to the potential performance gains, many tools favor \PDEVS over \DEVS in their implementation.
Some stick to the elegance of the original \DEVS formalism, despite the performance hit.

\subsection{Dynamic Structure DEVS}
Another shortcoming of the \DEVS formalism, also present in \PDEVS, is the lack of dynamic structure.
Some systems inherently require a dynamic structure to create, modify, or delete models during the simulation.
While possible in \DEVS formalism by defining the superset of all possible configurations and activating only one of them, this has high accidental complexity, and performance suffers.
Furthermore, systems might grow extremely big, making it practically impossible to create all possible configurations statically.

To counter these issues, \DSDEVS~\cite{DSDEVS} was devised as an extension of \DEVS.
In \DSDEVS, the model configuration is seen as a part of the state, making it modifiable during simulation.
Since the coupled model has no state of its own, a \textit{network executive} is added, which manages the structural state of a specific scope.
In a separate phase, models can send events to the network executive to request a structural change.

This proposed extension is, however, only a mathematical model as to how it can be made possible.
Similar to previous formalisms, an abstract simulator~\cite{DSDEVS_sim} is provided that is structured that way.
Real implementations, however, are free to implement this however they want.
The network executive might therefore not even exist in the implementation, with all structure changing messages being intercepted in the implementation.

Even though dynamic structure now becomes possible in \DEVS models, this formalism is not well suited to handle a huge amount of changes.
The work to be done for a change, both for the user and the implementation, is just too time-consuming to execute frequently.
But even while highly dynamic models are not ideally suited, infrequent structural changes become very possible.

\subsection{Cell-DEVS}
Another variant of \DEVS presented here is the \CELLDEVS formalism.
Despite the elegance of the \DEVS formalism, it is still difficult to use it in a variety of situations, specifically in the context of cellular models.
\CELL~\cite{CellularAutomata} are a popular choice in the domain of cellular models, but contrary to the discrete-event nature of \DEVS, \CELL is discrete-time based.
While discrete-time is a good match with most models in the problem domain of cellular automata, some models would profit from a discrete-event basis.
While not frequently a problem, cellular models become restricted to the granularity of the time step, resulting in low performance when the time step is not a good match with the model's transition times.

\CELLDEVS was introduced as a combination of \DEVS and \CELL, combining the best of both worlds.
Model specification is similar to \CELL models, but the underlying formalism used for simulation is actually \DEVS.
Due to this change, models gain more control over the simulation time.
Furthermore, cellular models can now be coupled to other, not necessarily cellular, \DEVS models.

\subsection{Other Variants}
Apart from the formalisms introduced here, many more variants exist that tackle very specific problems in \DEVS.
We do not have the space here to discuss all of them, though we wish to provide some pointers to some other useful extensions.
Examples are other solutions to the dynamic structure problem (\textsf{DynDEVS}~\cite{DynDEVS}), restrictions to make \DEVS models analyzable (\textsf{FD-DEVS}~\cite{FDDEVS}), and extensions to allow for non-determinism (\textsf{Fuzzy DEVS}~\cite{FuzzyDEVS}).
Many of the previously proposed formalisms also have augmented themselves with the changes made to \PDEVS, resulting in a parallel version of \DSDEVS~\cite{DSDEVS_article} and \CELLDEVS~\cite{ParallelCellDEVS}.

\ifnotreport
    \reflection{What does it mean for there to be extensions to \DEVS, while we previously stated that \DEVS can be seen as a simulation assembly language?}
\fi

\section{Summary}
\label{sec:conclusions}
In this chapter, we briefly presented the core ideas behind \DEVS, a popular formalism for the modelling of complex dynamic systems using a discrete-event abstraction.
\DEVS is primarily used for the simulation of queueing networks, of which an example was given, and performance models.
It is most applicable for the modelling of discrete event systems with component-based modularity.
It can, however, be used much more generally as a simulation assembly language, or as a theoretical foundation for these formalisms.

Future learning directions on \DEVS can be found in the \textsc{Further Reading} section, which provides a list of relevant extensions on \DEVS, as well as mentions of some of the problems currently being faced in \DEVS.

\ifnotreport
    \input further_reading.tex
    \input questions.tex
\fi

\section*{Acknowledgement}
This work was partly funded with a PhD fellowship grant from the Research Foundation - Flanders (FWO).
Partial support by the Flanders Make strategic research centre for the manufacturing industry
is also gratefully acknowledged.

\bibliographystyle{plain}
\bibliography{bibliography}

\begin{thebibliography}{10}

\bibitem{DSDEVS}
Fernando~J. Barros.
\newblock Dynamic structure discrete event system specification: a new
  formalism for dynamic structure modeling and simulation.
\newblock In {\em Proceedings of the 1995 Winter Simulation Multiconference},
  pages 781--785, 1995.

\bibitem{DSDEVS_article}
Fernando~J. Barros.
\newblock Modeling formalisms for dynamic structure systems.
\newblock {\em ACM Transactions on Modeling and Computer Simulation},
  7:501--515, 1997.

\bibitem{DSDEVS_sim}
Fernando~J. Barros.
\newblock Abstract simulators for the {DSDE} formalism.
\newblock In {\em Proceedings of the 1998 Winter simulation Multiconference},
  pages 407--412, 1998.

\bibitem{Chen}
Bin Chen and Hans Vangheluwe.
\newblock Symbolic flattening of {DEVS} models.
\newblock In {\em Proceedings of the 2010 Summer Simulation Multiconference},
  pages 209--218, 2010.

\bibitem{ParallelDEVS}
Alex Chung~Hen Chow and Bernard~P. Zeigler.
\newblock {Parallel DEVS}: a parallel, hierarchical, modular, modeling
  formalism.
\newblock In {\em Proceedings of the 1994 Winter Simulation Multiconference},
  pages 716--722, 1994.

\bibitem{ParallelDEVSAbsSim}
Alex Chung~Hen Chow, Bernard~P. Zeigler, and Doo~Hwan Kim.
\newblock Abstract simulator for the parallel {DEVS} formalism.
\newblock In {\em AI, Simulation, and Planning in High Autonomy Systems}, pages
  157--163, 1994.

\bibitem{Rhapsody}
David Harel and Hillel Kugler.
\newblock The rhapsody semantics of statecharts (or, on the executable core of
  the {UML}).
\newblock In {\em Integration of Software Specification Techniques for
  Applications in Engineering}, volume 3147 of {\em Lecture Notes in Computer
  Science}, pages 325--354. Springer Berlin Heidelberg, 2004.

\bibitem{Statemate}
David Harel and Amnon Naamad.
\newblock The {STATEMATE} semantics of {Statecharts}.
\newblock {\em ACM Transactions on Software Engineering Methodology},
  5(4):293--333, 1996.

\bibitem{SequentialPDEVS}
Jan Himmelspach and Adelinde~M. Uhrmacher.
\newblock Sequential processing of {PDEVS} models.
\newblock In {\em Proceedings of the 3rd European Modeling \& Simulation
  Symposium}, pages 239--244, 2006.

\bibitem{FDDEVS}
Moon-Ho Hwang.
\newblock Generating finite-state global behavior of reconfigurable automation
  systems: {DEVS} approach.
\newblock In {\em Proceedings of the International Conference on Automation
  Science and Engineering}, pages 254 -- 260, 2005.

\bibitem{FuzzyDEVS}
Yi~Wan Kwon, Hyu~Chan Park, Sung~Hoon Jung, and Tag~Gon Kim.
\newblock Fuzzy-{DEVS} formalism: concepts, realization and application.
\newblock In {\em Proceedings of AI, Simulation and Planning in High Autonomy
  Systems}, pages 227 -- 234, 1996.

\bibitem{BuildingSimulator}
James~J. Nutaro.
\newblock {\em Building Software for Simulation: Theory and Algorithms, with
  Applications in {C++}}.
\newblock Wiley, 1st edition, 2010.

\bibitem{standardization}
Hessam~S. Sarjoughian and Yu~Chen.
\newblock Standardizing {DEVS} models: an endogenous standpoint.
\newblock In {\em Proceedings of the 2011 Spring Simulation Multiconference},
  pages 266--273, 2011.

\bibitem{ParallelCellDEVS}
Alejandro Troccoli and Gabriel Wainer.
\newblock Implementing {Parallel Cell-DEVS}.
\newblock In {\em Proceedings of the 2003 Spring Simulation Symposium}, pages
  273--280, 2003.

\bibitem{DynDEVS}
Adelinde~M. Uhrmacher.
\newblock Dynamic structures in modeling and simulation: a reflective approach.
\newblock {\em ACM Transactions on Modeling and Computer Simulation},
  11:206--232, 2001.

\bibitem{PythonPDEVS1}
Yentl Van~Tendeloo and Hans Vangheluwe.
\newblock The modular architecture of the {Python(P)DEVS} simulation kernel.
\newblock In {\em Proceedings of the 2014 Spring Simulation Multiconference},
  pages 387--392, 2014.

\bibitem{JDF}
Yentl Van~Tendeloo and Hans Vangheluwe.
\newblock An overview of {PythonPDEVS}.
\newblock In Collectif~Workshop RED, editor, {\em JDF 2016 -- Les Journ\'ees
  DEVS Francophones -- Th\'eorie et Applications}, pages 59--66, 2016.

\bibitem{DEVSSurvey}
Yentl Van~Tendeloo and Hans Vangheluwe.
\newblock An evaluation of {DEVS} simulation tools.
\newblock {\em SIMULATION}, 93(2):103--121, 2017.

\bibitem{InitializedDEVS}
Yentl Van~Tendeloo and Hans Vangheluwe.
\newblock Extending the {DEVS} formalism with initialization information.
\newblock {\em ArXiv e-prints}, 2018.

\bibitem{DEVSbase}
Hans Vangheluwe.
\newblock {DEVS as a common denominator for multi-formalism hybrid systems
  modelling}.
\newblock In {\em IEEE International Symposium on Computer-Aided Control System
  Design}, pages 129--134, 2000.

\bibitem{PractitionersApproach}
Gabriel~A. Wainer.
\newblock {\em Discrete-Event Modeling and Simulation: A Practitioner's
  Approach}.
\newblock CRC Press, 1st edition, 2009.

\bibitem{CellularAutomata}
Stephen Wolfram.
\newblock Statistical mechanics of cellular automata.
\newblock {\em Reviews of Modern Physics}, 55(3):601 -- 644, 1983.

\bibitem{ClassicDEVS}
Bernard~P. Zeigler, Herbert Praehofer, and Tag~Gon Kim.
\newblock {\em Theory of Modeling and Simulation}.
\newblock Academic Press, 2nd edition, 2000.

\end{thebibliography}

\end{document}